\documentclass[%
 aip,
 amsmath,amssymb,
preprint,%
]{revtex4-1}

\usepackage{lineno,hyperref}
\modulolinenumbers[5]
\usepackage{amsmath}
\usepackage{amsfonts}
\usepackage{amssymb}
\usepackage{fixltx2e}
\usepackage{graphicx,hyperref}
\usepackage{amssymb}
\usepackage{multirow}
\usepackage[dvipsnames]{xcolor}
\usepackage{xcolor,mdframed}
\usepackage{mathptmx}
\usepackage{listings}
\usepackage{float}
\usepackage{physics}
\usepackage{diagbox}
\usepackage{booktabs, tabularx}
\newcolumntype{C}{>{\centering\arraybackslash}X}
\makeatletter
\DeclareRobustCommand{\tsusb}[2]{{%
  \m@th\ensuremath{%
    ^{\mbox{\fontsize\sf@size\z@#1}}%
    _{\mbox{\fontsize\sf@size\z@#2}}%
  }%
}}
\makeatother
\newcommand{\tsb}{\textsubscript}
\newcommand{\tsu}{\textsuperscript}

\newcommand{\be}{\begin {equation}}

\newcommand{\ee}{\end {equation}}
\newcommand{\beqa}{\begin {eqnarray}}
\newcommand{\eeqa}{\end {eqnarray}}

\hypersetup{
    colorlinks,%
    citecolor=blue,%
    filecolor=blue,%
    linkcolor=blue,%
    urlcolor=blue
}

\usepackage{graphicx}
\usepackage{dcolumn}
\usepackage{bm}

\usepackage[utf8]{inputenc}
\usepackage[T1]{fontenc}
\usepackage{mathptmx}
\usepackage{etoolbox}

\begin{document}
\title {Detailed investigation on x-ray emission from laser driven high-Z foils in a wide intensity range : role of conversion layer and reemission zone}

\author{Gaurav Mishra}
\email{Author to whom correspondence should be adressed:gauravm@barc.gov.in}
\author{Karabi Ghosh}
\affiliation{Theoretical Physics Section, Bhabha Atomic Research Centre, Mumbai-400085, India}



\begin{abstract}
Detailed radiation hydrodynamic simulations are carried out to investigate x-ray emission process in four high-Z planar targets namely, tungsten (W), gold (Au), lead (Pb) and uranium (U) irradiated by 1 ns, 351 nm flat top laser pulses. A thorough zoning analysis is performed for all laser driven high-Z foils over a wide intensity range of $10^{12}-10^{15}$ W/cm\tsu{2} with appropriately chosen photon energy range and recombination parameter. The resulting variation of conversion efficiency over the full intensity range exhibits an optimum for all materials which is explained by considering the characteristic emission contributions from two different regions of laser irradiated plasma, namely, conversion layer and remission zone. A new generalized single scaling relation  based upon smooth broken power law is proposed for conversion efficiency variation along with the separate determination ($\eta_{S}$, $\eta_{M}$) in soft and hard/M-band x-ray regions. It has been observed that $\eta_{S}$ for Pb and W always lies in between that for Au and U for intensities smaller than $\sim 3\times 10^{13}$ W/cm\tsu{2}. On further increase in intensity, $\eta_{S}$ is observed to be maximum for Au and U whereas it is minimum for W. Significant contribution to M-band conversion efficiencies is observed in all elements for intensities higher than $\sim 2\times 10^{13}$ W/cm\tsu{2} with maximum and minimum values attained by W and U, respectively. The results are explained by considering the contributions from the emission coefficients of all materials in both conversion layer and reemission zone up to corresponding photon cut-off energies at different laser intensities.
\end{abstract}
\maketitle

\section{Introduction}

X-ray emission from laser produced plasmas (LPP) has attracted extensive interest due to its wide ranging applications in many fields, such as indirect drive inertial confinement fusion (ICF) \cite{lindl1995development,lindl2004physics,betti2016inertial}, high energy density physics \cite{benuzzi2006laser,barrios2013backlighter}, laboratory astrophysics \cite{remington2005high}, extreme ultraviolet (XUV) soft x-ray lasers \cite{tallents2003physics}, x-ray contact microscopy \cite{batani2002use} of biological samples and advanced lithography .\cite{freeman2011enhancements} All of these applications depend on the effectiveness of LPP which in turn is characterized by its intensity that is incident on the target under investigation. The main parameters controlling intensity of LPP are the drive laser energy, pulse duration and laser to x-ray conversion efficiency ($\eta$). Various efforts are put forward to enhance the conversion efficiency through an appropriate choice of laser and target parameters. In the context of indirect drive ICF scheme, use of short wavelength laser \cite{mead1983laser,nishimura1983radiation,goldman1986scaling,goldstone1987dynamics,mead1988modeling,nishimura1991x}, high-Z targets \cite {mochizuki1986atomic,popil1987measurement}, long pulse duration \cite{ze1989observation,davis2018soft}, mixture/cocktail \cite{nishimura1993x,orzechowski1996rosseland,chakera2003dependence,chaurasia2008x,dong2014enhancedJAP,dong2014enhancedPL} are proposed to enhance the conversion efficiency in the range of multi-eV to multi-keV. Recently, various other methods like CH foam coated gold targets \cite{xu2011beneficial}, low density high-Z foam target \cite{shang2013enhancement,shang2016experimental}, double foil (DF) target \cite{ge2015enhancement}, counter propagating (CP) laser beams irradiation \cite{ZhaoCP2017}, sandwiched target \cite{li2010study,haan2011point,li2021optimization}, multi-layer target \cite{ZhengML2021} and DF target irradiated with CP laser beams \cite{YuanDFCP2021} are considered to increase the conversion efficiency. In the field of multi-keV emission, different types of target geometries, including pre-exploded metallic thin foils \cite{girard2005multi,babonneau2008efficient}, high-Z liner \cite{girard2009titanium}, mid-Z or high-Z doped aerogels \cite{PhysRevLett.92.165005,fournier2009absolute}, etc. are investigated for more efficient x-ray emission. For the development of next generation lithography systems, role of various parameters like laser pulse duration, laser wavelength, target geometry, initial target density etc. \cite{freeman2011enhancements,ueno2007enhancement,higashiguchi2011extreme} are explored to maximise the conversion efficiency in extreme ultraviolet (EUV) region. 

Traditionally, various computational studies have been performed to understand the process of x-ray emission from laser irradiated high-Z materials \cite{mead1983laser,mead1988modeling,nishimura1983radiation,nishimura1991x,dewald2008x,olson2012x}. To gain proper physical insight of the underlying process, significant amount of theoretical work was undertaken by Sigel \emph{et al.} in 1990s \cite{sigel1990conversion}. The important outcome of the study was that the x-ray emitting high-Z target can be subdivided into three main regions, namely, 1) conversion layer (CL) with high temperature low density region towards laser irradiation side 2) reemission zone (RZ) with moderate density and temperature region and 3) shock wave (SW) with low temperature high density region.  With the help of dimensional analysis, separate scaling relations were obtained for various parameters like density, temperature etc. in RZ and CL. This theoretical model was further confirmed by numerical simulations \cite{eidmann1990conversion} of gold planar foils driven by sine square laser pulse of 0.3 ns. Another simple model of x-ray conversion was put forward by Guskov et \emph{al.} in planar geometry on the basis of ``average stationary corona'' model  \cite{gus1992simple}. They obtained order-of-magnitude explicit scaling laws between x-ray conversion efficiency and laser intensity in a limited range of $10^{14}-10^{15}$ W/cm\tsu{2}.  

In the case of indirect drive fusion, laser light shining on the inner side of a cylindrical cavity known as hohlraum gets partially converted into x-rays. Mutiple absorption and reemission of x-rays create a nearly isotropic radiation field that further compresses the fusion pellet and leads to initiation and burn of fusion reaction inside the pellet. High-Z materials like gold, depleted uranium or cocktail targets made of Au and U are preferred as hohlraum wall materials due to their higher wall albedo \cite{jones2007proof,schein2007demonstration,dewald2018first}. However, intense laser irradiation ($\sim 10^{14}$ W/cm\tsu{2}) of hohlraum wall also leads to emission of hard x-rays in keV regime, so called ``M-band'', along with soft x-rays.  \cite{olson2004shock,robey2005experimental} Soft x-rays lead to uniform compression of fusion pellet whereas hard x-rays are responsible for preheating of pellet and degrade the implosion characteristics \cite{olson2003preheat,cheng2016effects,zhang2019study}. Recently, we have investigated various high-Z elements (W, Au, Pb, U) and their mixtures in terms of wall albedo/wall loss and other ablation characteristics for a wide range of temperature drives \cite{mishra2018wide,mishra2019investigation}. Choice of Pb and W is motivated by the fact that Pb is a potential alternative of Au in LIFE based power reactors and W has the possibility of being used in tokamaks \cite{ross2013lead,safronova2009excitation}. The wall loss scaling relations obtained in the study are further used in hohlraum source-sink model \cite{rosen1996science} to determine the temperature in hohlraum with same conversion efficiency ($\eta = 0.7 t^{0.12}$, t is time in ns) \cite{lindl1995development,lindl2004physics} for all high-Z materials. Previously, the properties of hot dense plasma composed of above mentioned materials were also explored on the basis of average ionization and M-emissivity (emissivity integrated over all energies above the lower edge of the M-band of Au).\cite{scott2011study} In the present study, we intend to investigate the feasibility of these high-Z elements in terms of x-ray emission characteristics. The earlier studies on x-ray emission from high-Z materials (Au, U and cocktails) in soft and M-band regions have been carried out by Dewald et \emph{al.} \cite{dewald2008x}. The experiments were performed with spherical targets in a convergent direct drive setup and results were analyzed with 2D radiation hydrodynamic (RHD) simulations using LASNEX with XSN opacity model. Our study uses computationally less expensive 1D RHD simulations coupled with NLTE atomic physics modelling to carry out detailed investigations on the x-ray emission process in laser driven targets.  In particular, we have evaluated the conversion efficiencies (integrated over time and photon energy) of all four high-Z planar foils irradiated by 1 ns flat top laser pulses in a wide intensity range of $10^{12}-10^{15}$ W/cm\tsu{2}. Our results of x-ray emission clearly demonstrate the importance of correct photon energy range, dielectronic recombination and zoning analysis employed for all materials at each intensity used in the simulation. It may be noted that the conversion efficiency results obtained for planar foils can not be directly used in hohlraum energy balance model as hohlraum conversion efficiencies are found to be greater than their planar  counterpart due to accumulation of blowoff energy and material in the hohlraum \cite{massen1993modeling,suter1996radiation}. Nevertheless, our study conducted in planar foils, provides an accurate comparative analysis of the performance of the four high-Z elements. Our results also reveal an optimum in conversion efficiency variation with intensity for all materials which can be explained by considering the characteristics emission contributions from CL and RZ. Instead of separate relations for different regions, we have proposed a generalized single conversion efficiency scaling relation with laser intensity for all high-Z materials, based upon smooth broken power law. As opposed to the previous studies, our generalized scaling directly connects the conversion efficiency with the experimentally controllable and direct parameter - laser intensity in a wide range. Therefore, it can be easily used in characterizing the LPP sources based upon different materials in a wide range of laser intensities. Additionally, we have also compared the materials in terms of separately determined soft and M-band conversion efficiencies at different laser intensities. These results are explained as follows 1) first, representative locations of CL and RZ are determined in terms of density and temperature at different laser intensities 2) emission coefficients are evaluated at these locations 3) photon cut-off energies are decided by ignoring the relatively less dominant contributions in front spectrum 4) and lastly, the materials are compared in terms of emission coefficients evaluated at representative locations of CL and RZ up to appropriately chosen photon cut-off energies. This paper is organised as follows. In Section \ref{SimuModel}, we describe the different simulation models employed in this study. Section \ref{valid} validates our opacity modelling and methodology of conversion efficiency evaluation with earlier published simulation and experimental results. Detailed findings of the present study are discussed in section \ref{resul}. Finally, important conclusions and directions of future work are summarized in section \ref{conclu}. 
\section{Simulation model}\label{SimuModel}
The x-ray emission studies are carried out with a one-dimensional RHD code \cite{mishra2018wide,mishra2019investigation,gupta1995angular,gupta2001effects} which is a modified version of code MULTI \cite{ramis1988multi} and hence is not discussed in great detail. The radiation transfer equation is solved in multi-group radiation diffusion approximation. Mass, momentum and energy conservation equations are solved in Lagrangian formulation \cite{zel2002physics} and a flux limiting factor of 0.03 is used for all simulations presented in this paper. The coupled RHD equations are solved in a fully implicit manner with the time splitting scheme. Sub-cycling of hydrodynamics is allowed for every time step of radiation transport. 

Equation of state (EOS) data for all the elements used in this study are obtained from in-house EOS code, TFEOS \cite{ray2006improved}, applicable for wide range of density  and temperature. NLTE opacity tables for all materials are generated from SNOP radiative opacity code \cite{eidmann1994radiation}. SNOP model is an explicit ion model based on screened hydrogenic approximation where  population densities of different ionic species are determined by solving steady state rate equation. In ionization balance equation, atomic processes like electronic ionization, collision and radiative recombination are accounted for by use of explicit coefficients whereas dielectronic recombination is treated via a parameter denoted by ``d''.  A total number of 3000 photon frequency points, with lower and upper boundaries at 1 eV and 5 keV, are used to calculate the opacities for all materials. Further, total 20 photon energy groups are employed to determine the Rosseland and Planck mean opacities. Material density ranges from $10^{-6}$ g/cc to $100$ g/cc while temperature varies from 1eV to 100 keV. 
\section{Results}\label{Mainresul}
In this section, we first validate our results of opacity modelling and  x-ray conversion efficiencies with the earlier published results. Thereafter, main results of this paper will be described.
\subsection{Validation and verification}\label{valid}
\begin{figure}
        \includegraphics[width=0.49\textwidth]{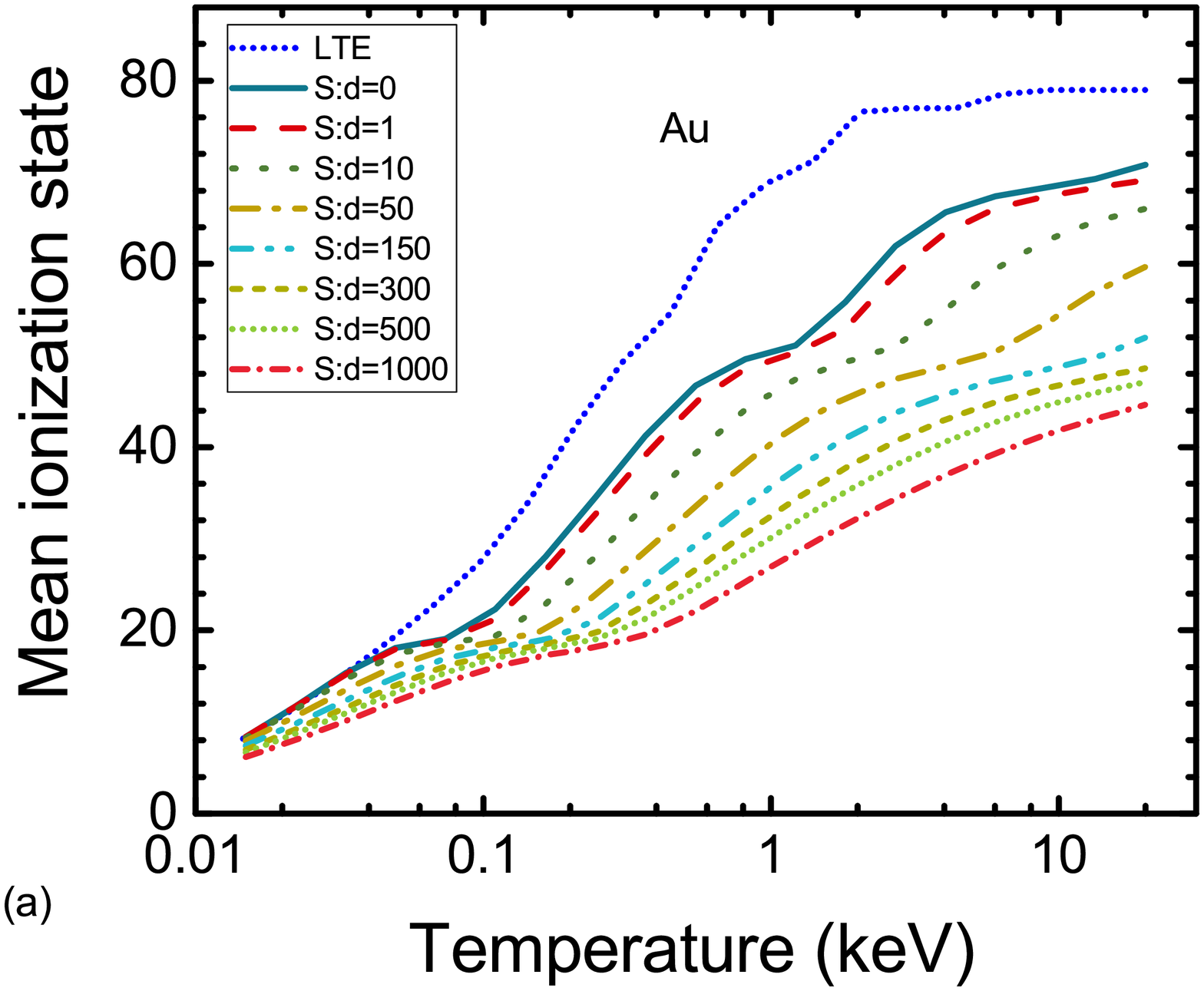}
        \includegraphics[width=0.49\textwidth]{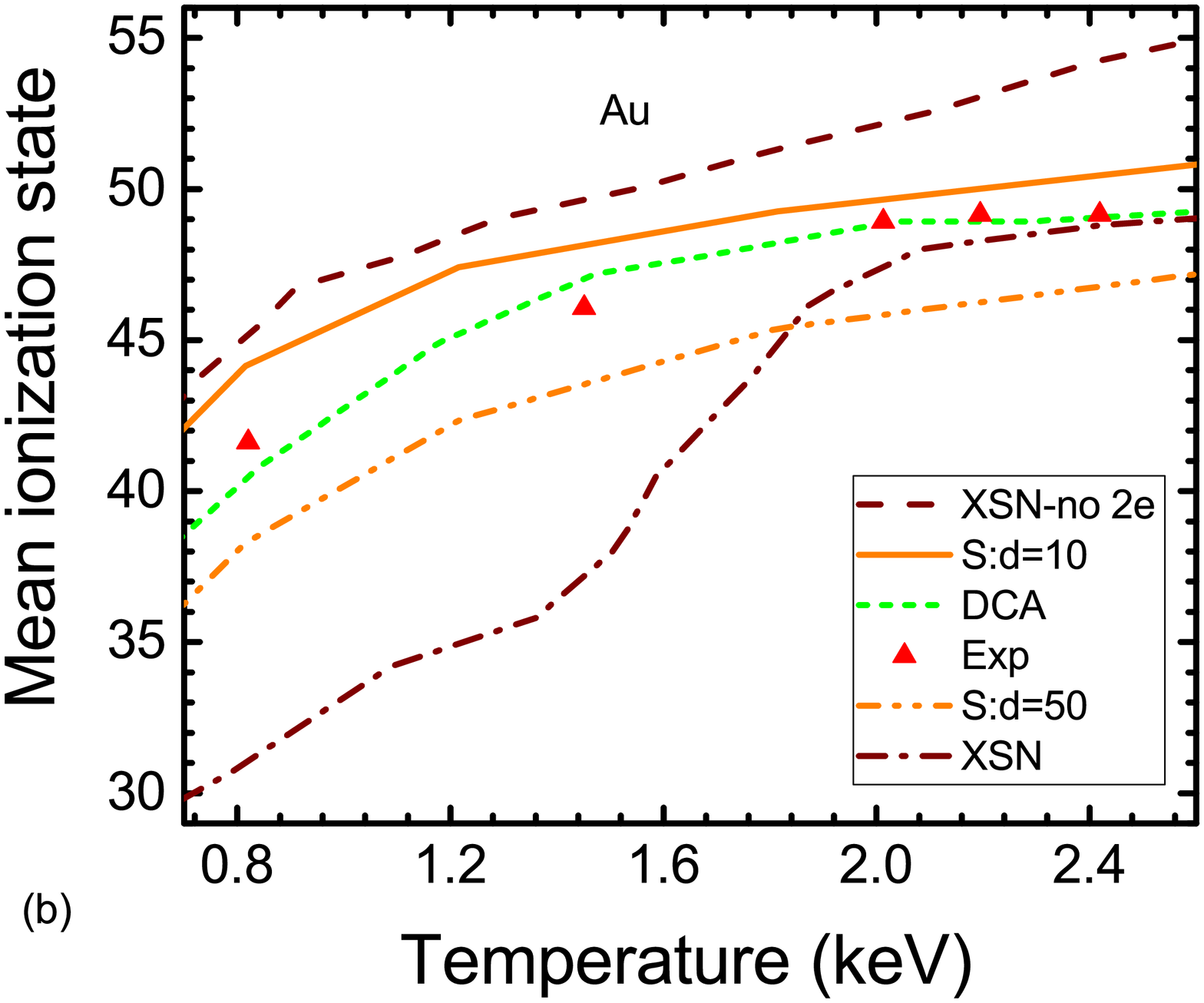}
        \includegraphics[width=0.49\textwidth]{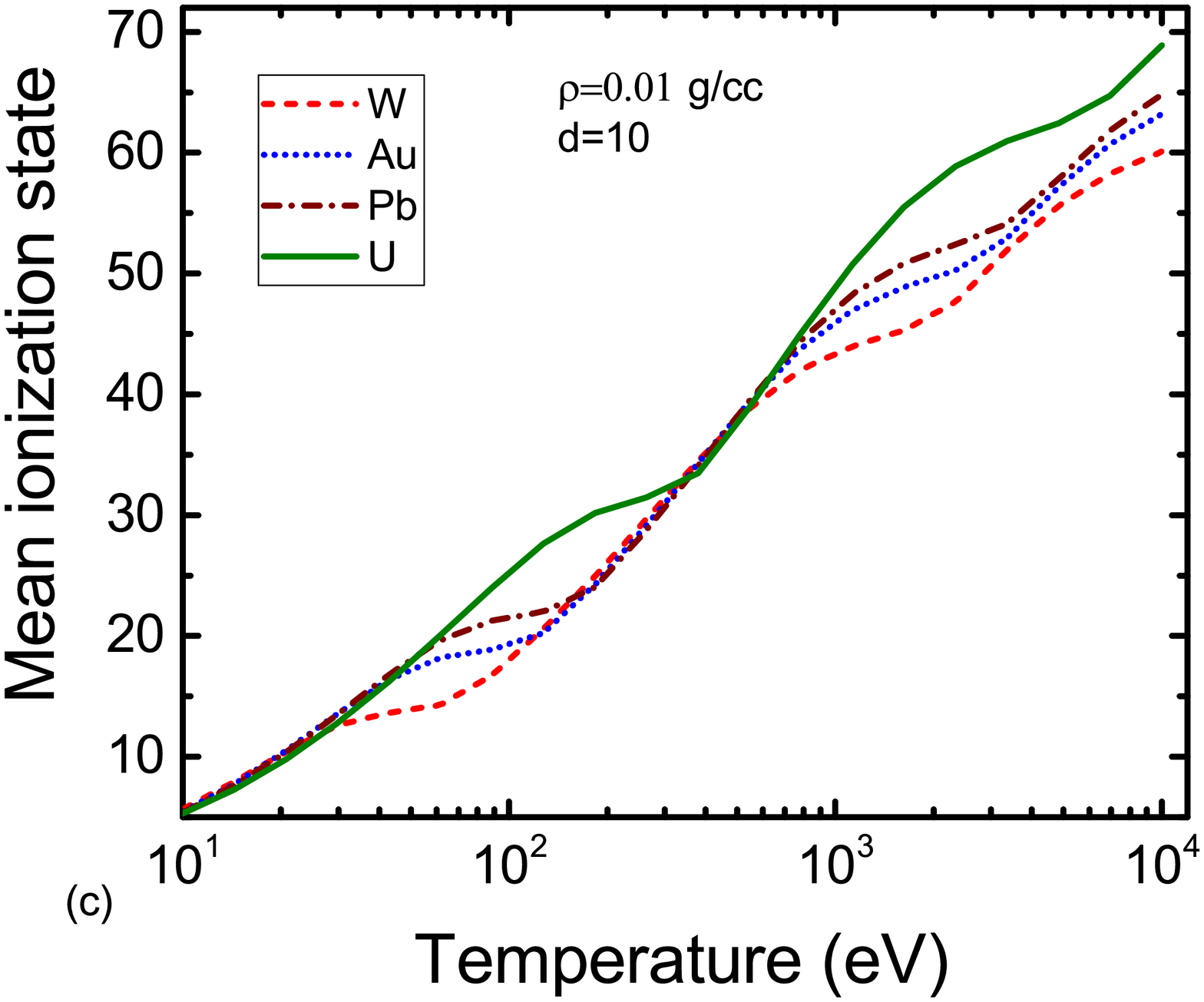}
\caption{(Colour online only) Comparison of mean ionization state evaluated (a) from LTE and SNOP (with different values of recombination parameter ``d'') opacity models  (b) from SNOP (d = 10 and 50), DCA, experiments, XSN-no 2e and XSN  for Au (c) from SNOP (d = 10) for all high-Z materials against temperature.\label{AuZbar}}
\end{figure}

In order to analyze the x-ray emission process in four high-Z planar targets, namely, W, Au, Pb and U using 1 ns flat top laser pulses, it is essential to first validate our results with those available in the published literature. The values of conversion efficiency obtained from RHD simulation depends on accurate modelling of various zones formed in the laser irradiated plasma which in turn depends upon the opacity model used. To check the accuracy of the opacities used in the present study, we have plotted mean ionization state ($\bar{Z}$) of gold with temperature in Fig. \ref{AuZbar} (a) for different values of \emph{d}, obtained from SNOP opacity model. 
This figure also includes $\bar{Z}$ from a more detailed LTE opacity code AALS which was previously developed \cite{mishra2018wide,mishra2019investigation}. We note that the use of NLTE opacity reduces $\bar{Z}$ compared to LTE opacity. To show the sensitivity of dielectronic recombination  on NLTE results, we have varied the value of d from 0 to 1000. We find that the mean ionization is significantly reduced from d = 0 to d = 1000. To ascertain the appropriate value of recombination parameter, we have compared $\bar{Z}$ from SNOP for d = 10 and d = 50 (S:d=10 and S:d=50) with other theoretical models (XSN, XSN without dielectronic recombination, DCA) and experiments in Fig. \ref{AuZbar} (b) for a smaller temperature range \cite{post1977steady,scott2010advances,scott2011study}. It is observed that average atom model such as XSN without 2e treatment overestimates mean ionization state whereas XSN with 2e treatment matches results at relatively high temperature but shows large disagreement at low temperature. The experimental results are well reproduced by use of DCA model that includes proper accounting of dielectronic recombination. We note that the mean ionization states evaluated from SNOP with d=10 are also close to experimental results. Therefore, we have used d = 10 for all simulation results presented in this study. Next, we present $\bar{Z}$ for all high-Z materials in temperature range of 10 eV to 10 keV in Fig. \ref{AuZbar} (c) at a density of 0.01 g/cc. Except for plateau related to energy shell structures, maximum and minimum mean ionization states are obtained for U and W, respectively, as the temperature is increased. 

\begin{figure}[] 
        \includegraphics[width=0.49\textwidth]{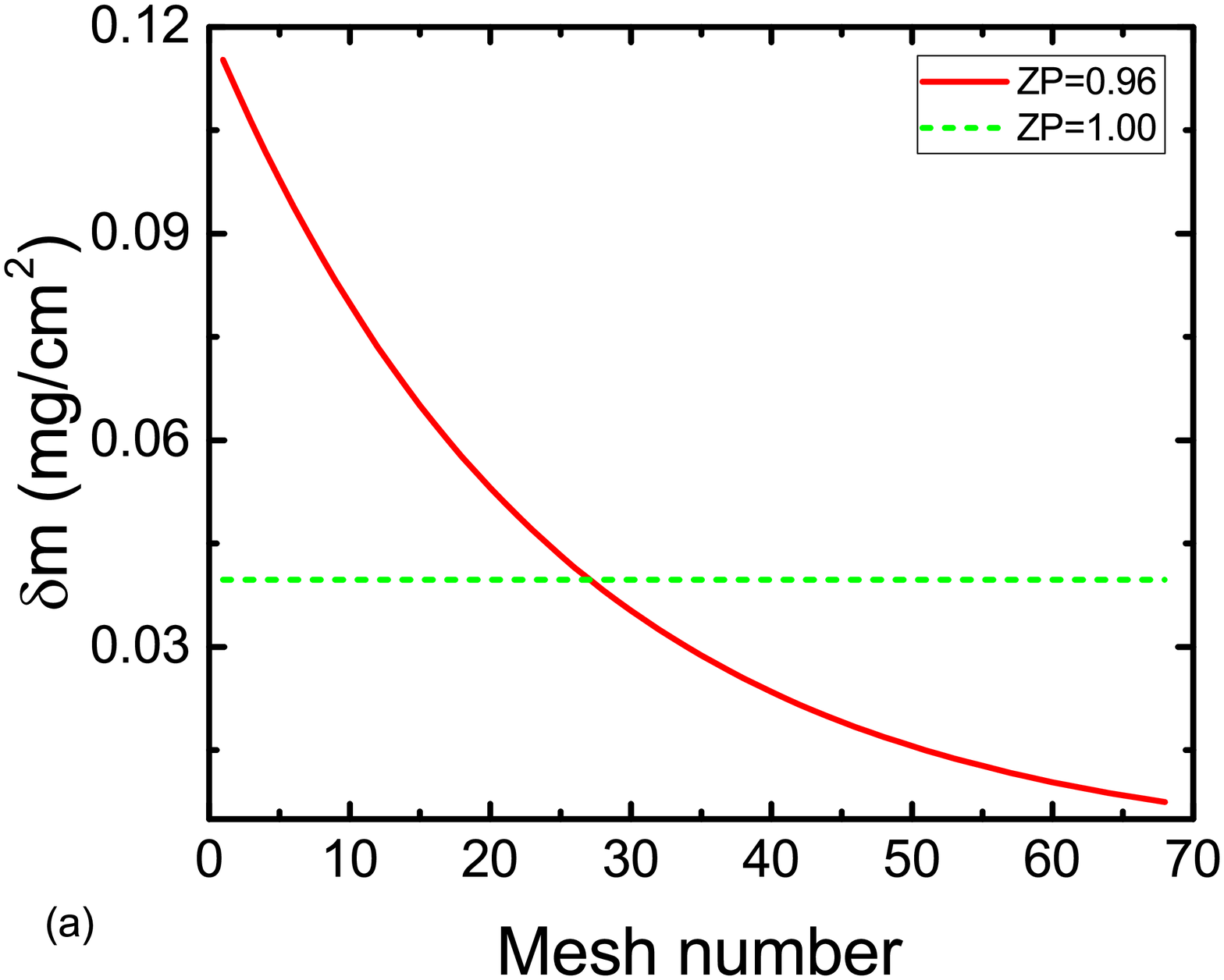}
        \includegraphics[width=0.49\linewidth]{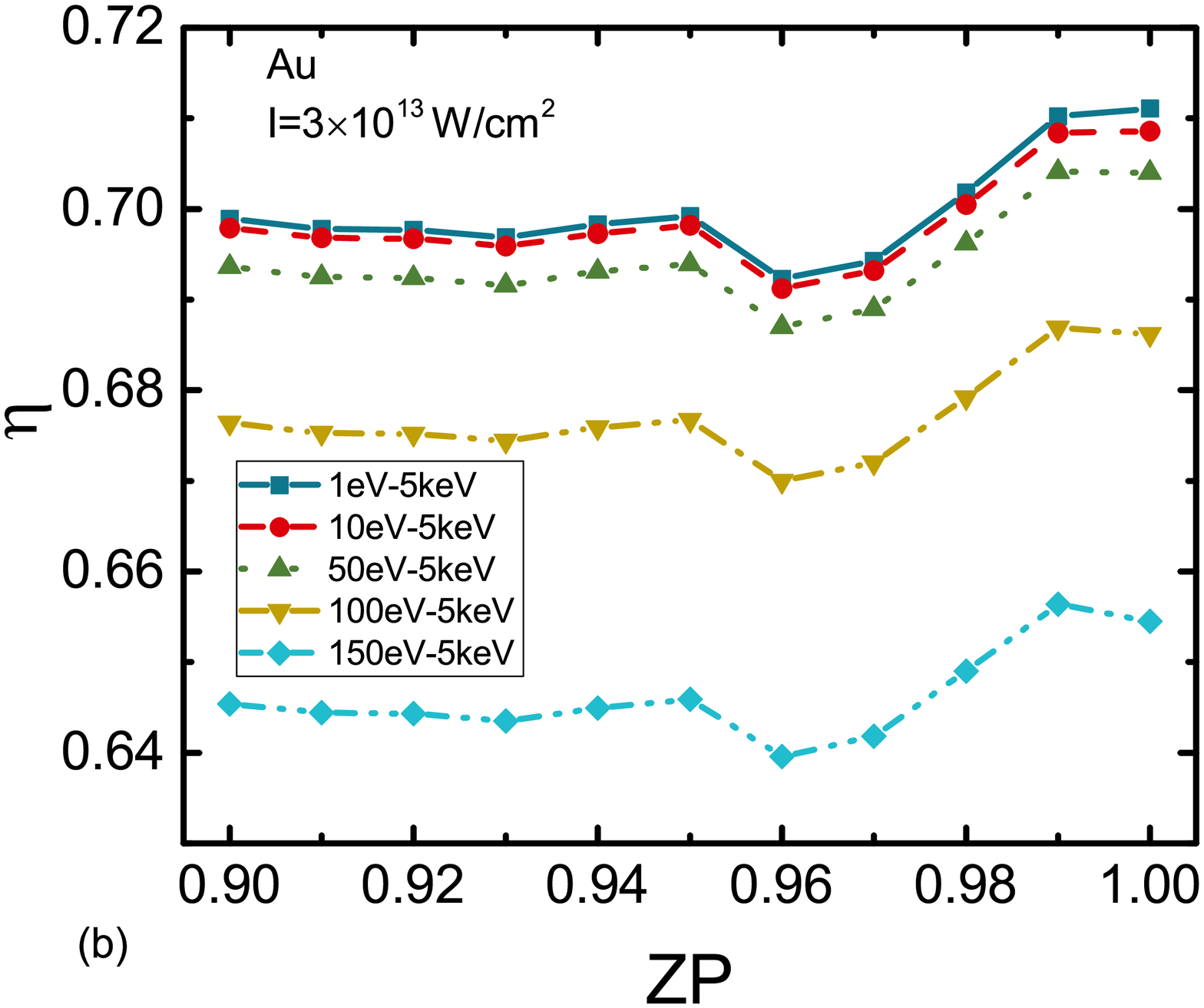}
\caption{(Colour online only) (a) Lagrangian mass width of different cells in a Au foil for two values of ZP, namely 0.96 and 1. (b) Variation of conversion efficiency with ZP in the same foil driven by sine square laser pulse of intensity $3\times 10^{13}$ W/cm\tsu{2}. The different lines correspond to $\eta$ evaluated using different photon energy ranges. Other laser and target parameters for the two plots are taken from Ref-1.\cite{eidmann1994radiation}}
\label{EtaSNOP}
\end{figure}

\begin{figure}[]
        \includegraphics[width=0.49\textwidth]{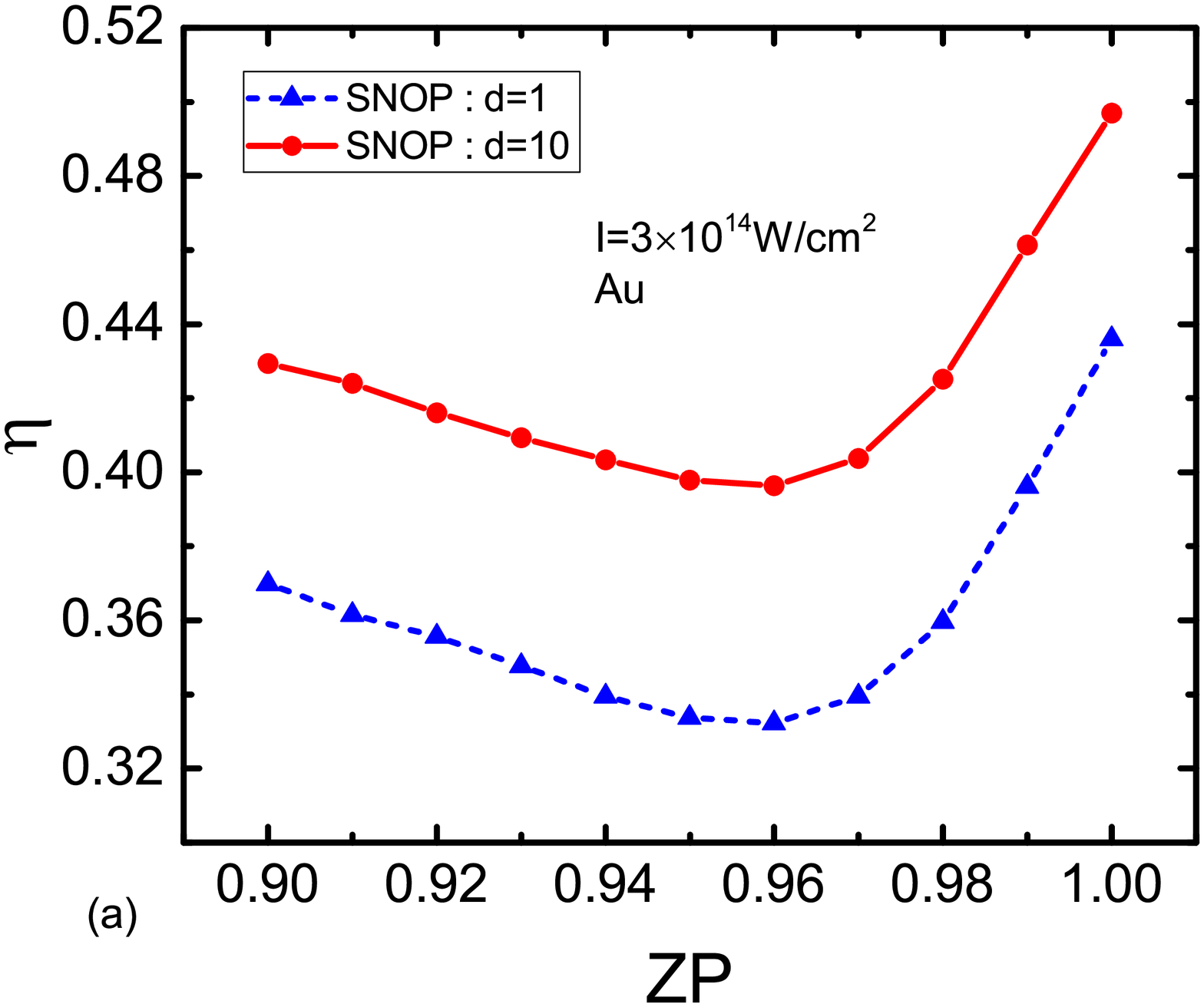}
        \includegraphics[width=0.49\textwidth]{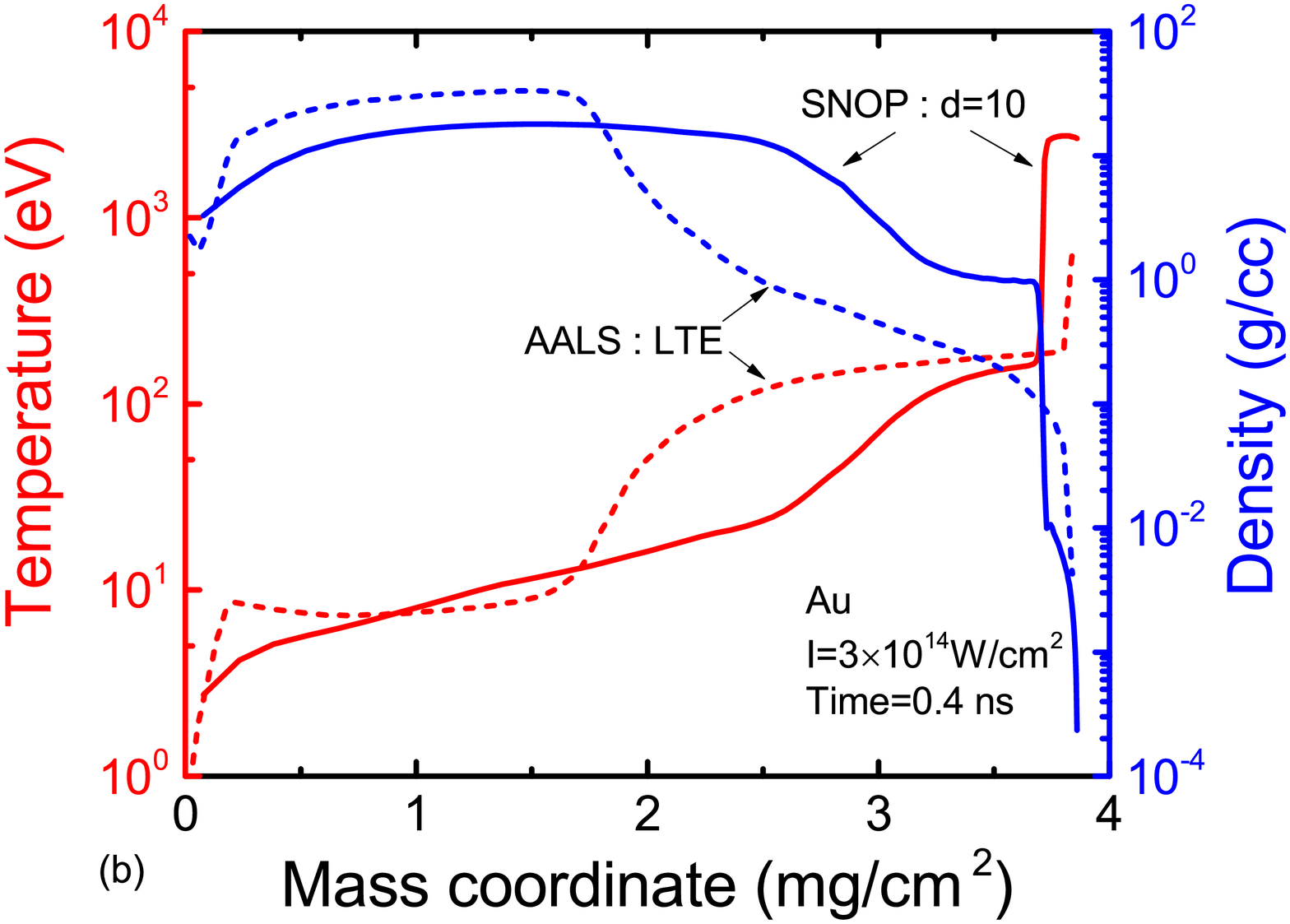}
\caption{(Colour online only) (a) Conversion efficiency for Au foil driven by flat top laser pulse of intensity $3\times 10^{14}$ W/cm\tsu{2} as a function of zone parameter. The laser and target parameters are taken from Ref. \cite{shang2016experimental}. The two different lines correspond to $\eta$ obtained from SNOP using different values of recombination parameters (d = 1 and d = 10). (b) Temperature (left) and density (right) variation as a function of mass coordinate for this set of laser and target parameters. The dashed and solid lines correspond to two different opacity models, namely AALS and SNOP with d = 10. \label{AuEtaShang}}
\end{figure}

 Now, we will validate our results of x-ray conversion efficiency with already published ones to verify our methodology of evaluation and fixing the choices of different parameters used for further analysis. In our studies, we have defined laser to x-ray conversion efficiency as total radiation energy (integrated over time and photon energy) emitted towards laser side, normalised to the absorbed laser energy. For comparison, we have selected two case studies : 1) Ref-1\cite{eidmann1994radiation} 2) Ref-2\cite{shang2016experimental}. Ref-1 describes the x-ray conversion process in gold foil of thickness 1.4 $\mu$m irradiated with sine-square laser pulse of FWHM 1 ns and intensity $3\times 10^{13}$ W/cm\tsu{2}. The total x-ray conversion efficiency was found to be around 0.64 at the end of the laser pulse. Our simulation results are found to be sensitive to mesh thickness of the foil. To resolve large gradients observed for hydrodynamic flow variables in CL, fine spatial meshing is essential close to laser side of foil whereas RZ does not pose such requirement. Instead of taking the same mesh size for all Lagrangian cells, nonuniform mesh widths are used by invoking zone parameter (ZP) \cite{ramis1988multi} in RHD simulations. The Lagrangian mesh widths ($\delta m$) of consecutive cells are related as, $\delta m_{i+1} = ZP\times \delta m_{i}$, where $i$ is mesh number starting from left. The value of ZP as unity refers to situation of uniform mesh width whereas its value greater or less than one dictates finer zoning towards left or right side, accordingly. For all calculations presented in this paper, laser is assumed to shine on target from right side so we expect ZP values smaller than one should be able to resolve the steep gradient observed towards laser irradiating right side. For demonstration purpose, we have plotted Lagrangian mass width of different cells in Fig. \ref{EtaSNOP} (a) for the above mentioned problem of laser driven gold foil. We observe the mesh width to be uniform for ZP = 1 whereas it reduces towards laser irradiation side for ZP = 0.96. We have carried out zoning studies for the given laser intensity and foil thickness. Similar kind of zoning studies were performed in previous studies too \cite{goldman1986scaling,mead1988modeling}. It may be noted that while obtaining Rosseland and Planck opacity, 20 photon energy groups were considered over the range of 1 eV to 5 keV. To accurately determine the photon energy range for x-ray emission, we have evaluated  conversion efficiencies integrated over different ranges and compared with the result of Ref-1. The results are shown in Fig. \ref{EtaSNOP} (b) against ZP for different photon energy ranges. 
 For all photon energy ranges, $\eta$ is found to be minimum for a ZP value of 0.96. Moreover, it turns out to be  around 0.6396 for a photon energy range of 150 eV - 5 keV and shows best agreement with the result of Ref-1 (0.64). Hence for all further simulation studies presented in this paper, we have determined $\eta$ in photon energy range of 150 eV to 5 keV. Next, we consider experimental and simulation studies conducted in Ref-2 where Au planar foils are irradiated with 1 ns flattop laser pulse of wavelength 351 nm. The experimentally measured $\eta$ was found to be around 0.414 at laser intensity of  $3\times 10^{14}$ W/cm\tsu{2} that matched with their simulations well. In our studies, we have used 1 ns flattop laser pulse with 100 ps rise and fall time. The simulated x-ray conversion efficiencies for photon energy range 150 eV to 5 keV are plotted against ZP for two different values of recombination parameter (d = 1, 10) in Fig. \ref{AuEtaShang} (a). Earlier, the effect of \emph{d} was shown in terms of mean ionization state. Now, we directly measure the change in $\eta$ due to use of different values of d. For both d = 1 and d = 10, we observe that $\eta$ shows a clear minimum at around 0.96. The sensitivity towards ZP is more pronounced as the laser intensity is one order higher in this case compared to that of Ref-1. For recombination parameter of d = 10, $\eta$ is 0.4970 for uniform mesh width (ZP = 1), whereas the same reduces to a minimum value of 0.3964 for non-uniform mesh width having ZP = 0.96. For d = 1, the conversion efficiencies are reduced from 0.4360 (ZP = 1) to a minimum value of 0.3322 for ZP = 0.96. The value of $\eta$ obtained for d = 10 and ZP = 0.96 is 0.3964 which is closer to that of Ref-2 (0.414). For this set of simulation parameters, the spatial profile of electron temperature and density are plotted against Lagrangian mass coordinate in Fig. \ref{AuEtaShang} (b) at a time of 0.4 ns. We clearly observe the three main regions obtained in laser irradiated high-Z plasma profile namely CL, RZ and SW. The temperature and density profiles obtained with  AALS\cite{mishra2018wide,mishra2019investigation} are also plotted in the same figure .   We observe the inadequacy of LTE opacities to correctly reproduce the variations of hydrodynamic flow variables. LTE opacities give a cooler corona temperature ($T{^{LTE}_{e}} = 677$ eV) compared to that obtained from NLTE opacities ($T{^{NLTE}_{e}} = 2.664$ keV) because in LTE, the specific emission being independent of the density, the maximum emission occurs at the CL where the temperature is the highest \cite{eidmann1990conversion}. Also, the RZ is extended compared to NLTE opacities. Further, relatively higher densities are obtained in CL with use of LTE opacities compared to NLTE opacities. As radiative and dielectronic recombinations are not considered, the radiation emission is far more effective with LTE physics. Thus, for LTE opacities, we obtain a very high value of $\eta$ (0.8448).
\subsection{Conversion efficiency of different elements}\label{resul}
\begin{figure}[]
        \includegraphics[width=0.49\textwidth]{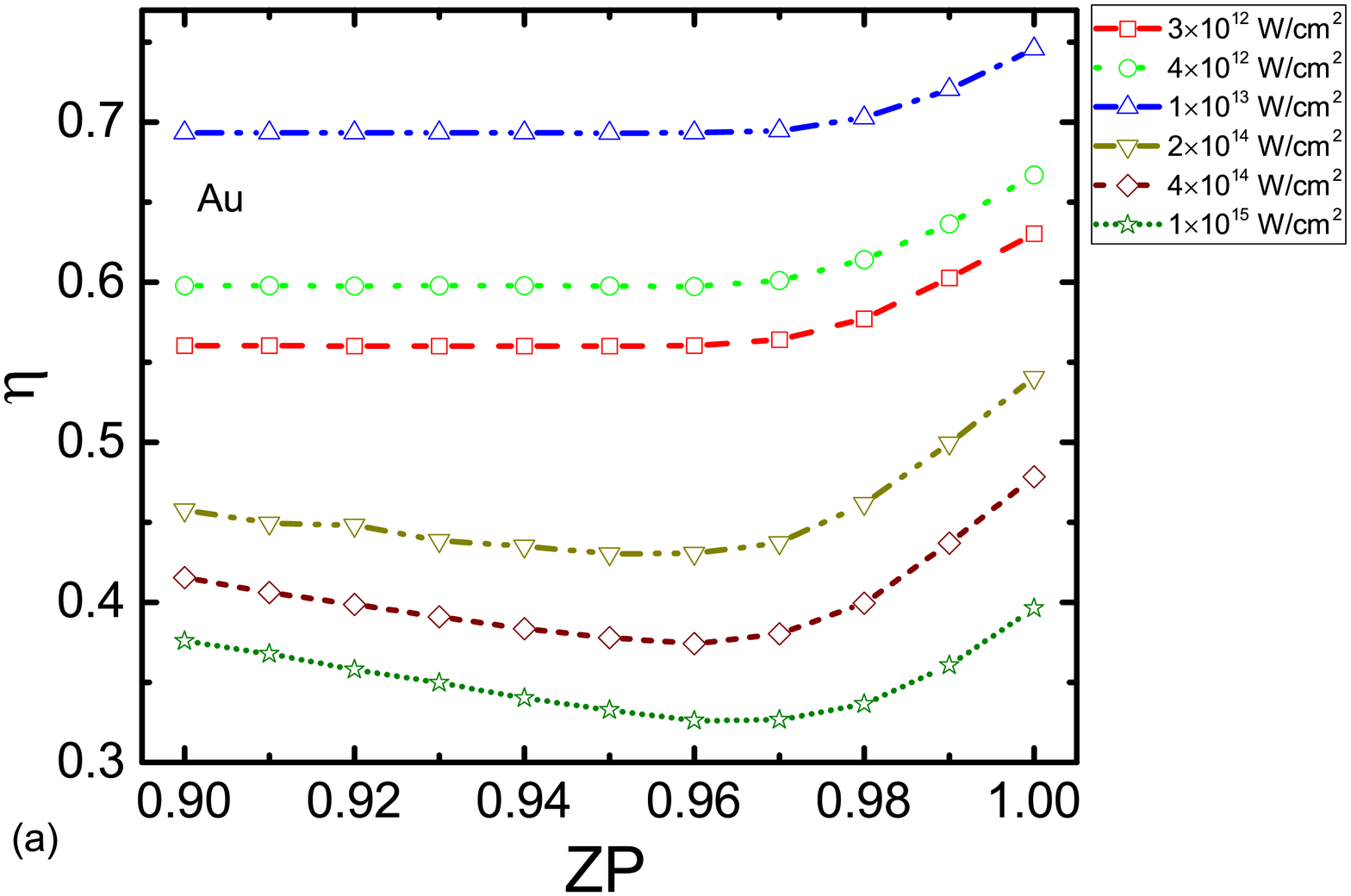}
        \includegraphics[width=0.49\textwidth]{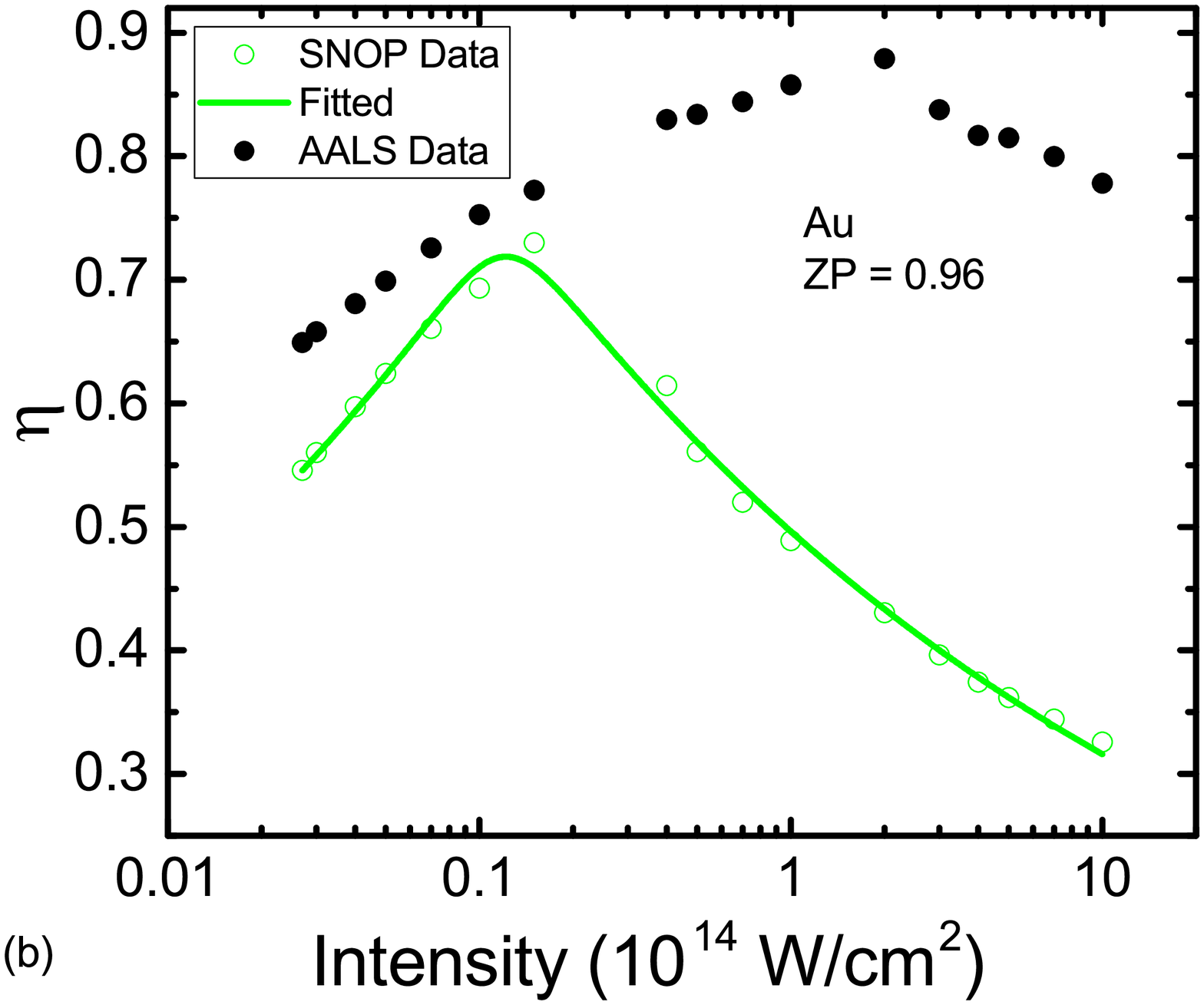}
\caption{(Colour online only) (a) Conversion efficiency of Au foil driven by flat top 1 ns laser pulse as a function of zoning parameter at different laser intensities. (b) $\eta$  data and fitted results from Eq.\ref{EqEtaFit} for ZP = 0.96 against laser intensity (in the unit of $10^{14}$ W/cm\tsu{2}). \label{AuEta}}
\end{figure}

In this subsection, we present RHD simulation results on x-ray emission in all high-Z laser driven planar targets. First, we describe the results of conversion efficiency variation with laser pulses of varying intensities in representative material gold. The planar foil of thickness 2 $\mu$m, corresponding to an areal mass density of 3.86 mg/cm\tsu{2}, is irradiated with 0.351 $\mu$m flat top laser of pulse duration 1 ns with 100 ps rise and fall time. The intensity of laser pulse is varied over a range of $10^{12}$ W/cm\tsu{2} to $10^{15}$ W/cm\tsu{2} whereas $\eta$ is determined at 1.6 ns for all laser intensities. Each foil is divided into 98 Lagrangian cells. Extensive zoning studies are performed for each laser intensity employed in the simulations. In Fig. \ref{AuEta} (a), we plot $\eta$  against ZP over the range of 0.9 to 1 for different laser intensities.
For lower laser intensities, values of $\eta$ first decrease from ZP = 1 and then converge after ZP = 0.96. At higher intensities, the sensitivity of $\eta$ increases with ZP as observed earlier in Ref-2. Moreover, $\eta$ first decreases as ZP is reduced from 1, showing that thinner meshes are required to adequately resolve steep temperature and density gradients encountered in CL. After reaching an optimum at ZP = 0.96, $\eta$ shows an upward trend on further reduction in ZP. This happens because the flow variables in plasma region are getting overly resolved on reducing ZP below 0.96. We note that $\eta$ either converges or shows an optimum value at 0.96 for all laser intensities used in the simulations. So, we have chosen ZP = 0.96 for all cases presented in the study. In Fig. \ref{AuEta} (b), we have presented  results of $\eta$  against laser intensity for ZP = 0.96. It may be noted that laser intensity in this plot is normalized in units of $10^{14}$ W/cm\tsu{2} so normalized values span the range of 0.01 to 10. We observe from the plot that $\eta$ first increases with intensity, passes through a maximum at an intensity of $\sim 10^{13}$ W/cm\tsu{2} and then decreases monotonously on further enhancement of laser intensity. For the sake of comparison, we have also shown conversion efficiencies obtained from AALS LTE opacities in Fig. \ref{AuEta} (b). We observe that LTE results over-predict $\eta$ for all laser intensities. Moreover, the difference is significant at higher intensities. Although LTE opacities also show a maximum in $\eta$, it is observed at a much higher intensity of $2\times10^{14}$ W/cm\tsu{2} and the fall at higher intensities is also not as sharp as obtained with SNOP NLTE opacities. This kind of characteristic behaviour of conversion efficiency with laser intensity has been observed in previous studies \cite{mead1983laser,nishimura1983radiation,goldstone1987dynamics}. By using dimensional analysis, Sigel et al. \cite{sigel1990conversion} investigated x-ray conversion process theoretically and predicted that with increasing laser flux, the conversion efficiency first increases, reaches a maximum, and then decreases. Moreover, they obtained scaling laws for density, ablated mass, temperature and relative hydroloss with source flux in RZ and with incident laser flux in CL using dimensional analysis. Further, the dependence of x-ray emission on intensity, wavelength and pulse duration of the laser was obtained by performing numerical studies on x-ray emission from gold foil driven by sine-squared pulse of 300 ps by Eidmann \emph {et al.} \cite{eidmann1990conversion}. Later, Guskov et \emph{al.} proposed order-of-magnitude explicit scaling laws for x-ray conversion efficiency with laser intensity in the limited range of $10^{14}-10^{15}$ W/cm\tsu{2} on the basis of ``average stationary corona'' model for short wavelength lasers  \cite{gus1992simple}. Based on these scaling relations, we have proposed the following smooth broken power law for conversion efficiency data against normalized laser intensities ($\tilde{I}$) applicable in the wide intensity range of $10^{12}-10^{15}$ W/cm\tsu{2},
\be
\eta=\eta_0 \left(\frac{\tilde{I}}{\tilde{I_0}}\right)^a\left[ 1+\left(\frac{\tilde{I}}{\tilde{I_0}}\right)^\frac{a+b}{c}\right]^{-c}
\label{EqEtaFit}
\ee 
where $\eta_0,\tilde{I_0},a,b,c$ are the fitting coefficients. Here, $\tilde{I_0}$ can be interpreted as break intensity. $\eta$ satisfies approximate power laws with the index values $a$ and $b$ for intensities $\tilde{I}\lesssim \tilde{I_0}$ and $\tilde{I}\gtrsim\tilde{I_0}$, respectively. The two power laws are smoothly joined close to $\tilde{I_0}$ where parameter $c$ sets the smoothness of the slope change. The form of the power law  given in Eq.\ref{EqEtaFit} can be further simplified to $\left(\eta=\eta_0\middle/\left[\left(\tilde{I}/\tilde{I_0}\right)^{-a/c}+\left(\tilde{I}/\tilde{I_0}\right)^{b/c}\right]^{c}\right)$. For Au, the values of the fitting coefficients are given in Table \ref{table:IntenFit} and the fitted results are shown with solid line in Fig. \ref{AuEta} (b).
\begin{table*}
\caption{Coefficients of power law fits of conversion efficiency from Eq. \ref{EqEtaFit} for different high-Z elements.\label{table:IntenFit}}
\begin{ruledtabular}
\begin{tabular}{ccccc}
                &W     &Au    &Pb    &U\\ \hline
$\eta_0$        &0.72  &0.75  &0.82  &1.15\\ 
$\tilde{I_0}$   &0.12  &0.12  &0.12  &0.12\\
$a$             &0.29  &0.22  &0.37  &0.40\\
$b$             &0.24  &0.20  &0.26  &0.31\\
$c$             &0.08  &0.06  &0.38  &0.97
\end{tabular}
\end{ruledtabular}
\end{table*}    
The simulation data points are found to lie on the fitted curve at lower and higher intensities. However, slight mismatch is observed close to the break intensity. As laser intensity is an experimental parameter that can be directly controlled, the above proposed scaling relation (Eq.\ref{EqEtaFit}) between conversion efficiency and laser intensity may prove to be useful in predicting the applicability of various LPP sources.
\subsubsection{Role of conversion layer and reemission zone}
\begin{figure}[]
        \includegraphics[width=0.49\textwidth]{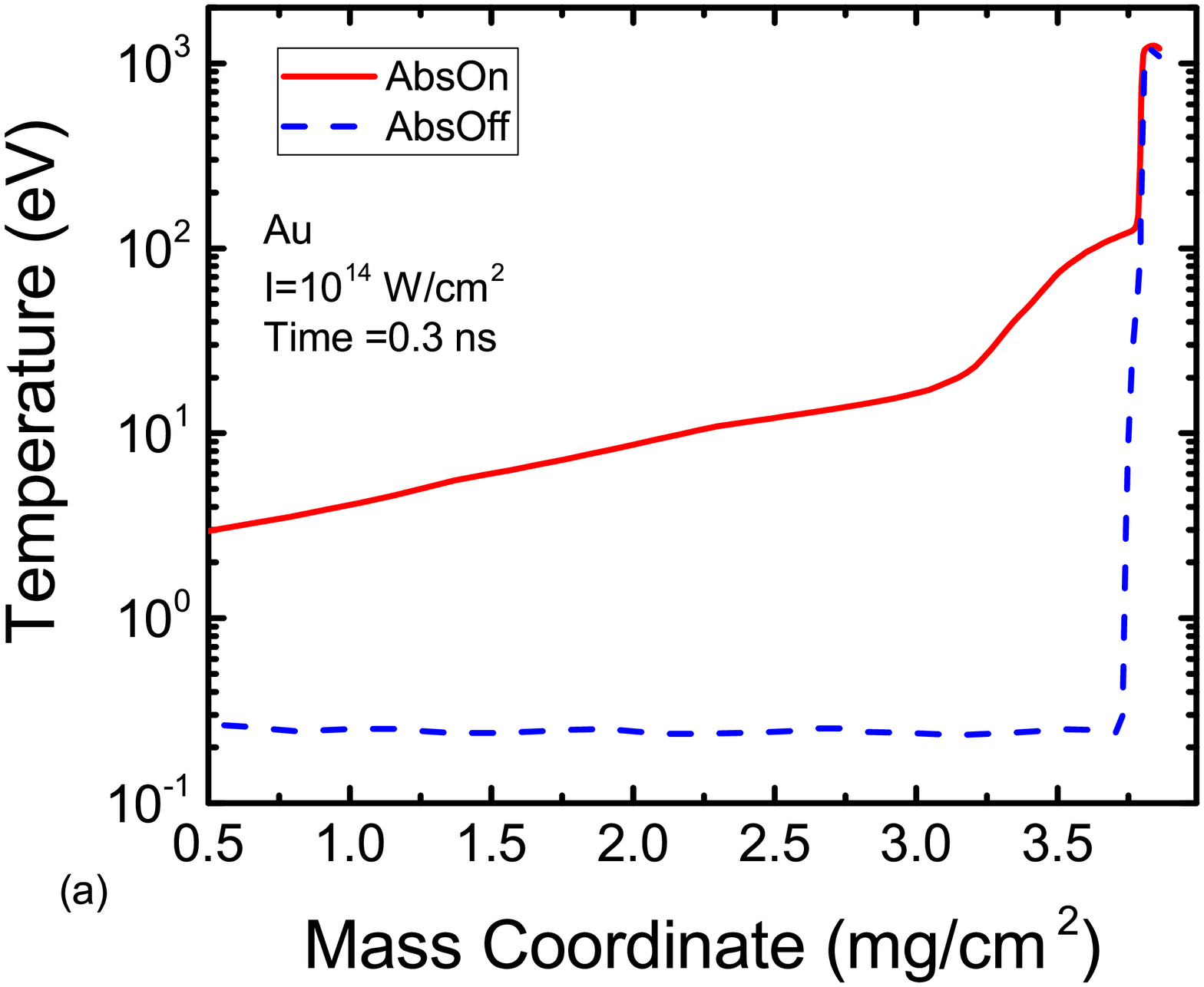}
        \includegraphics[width=0.49\textwidth]{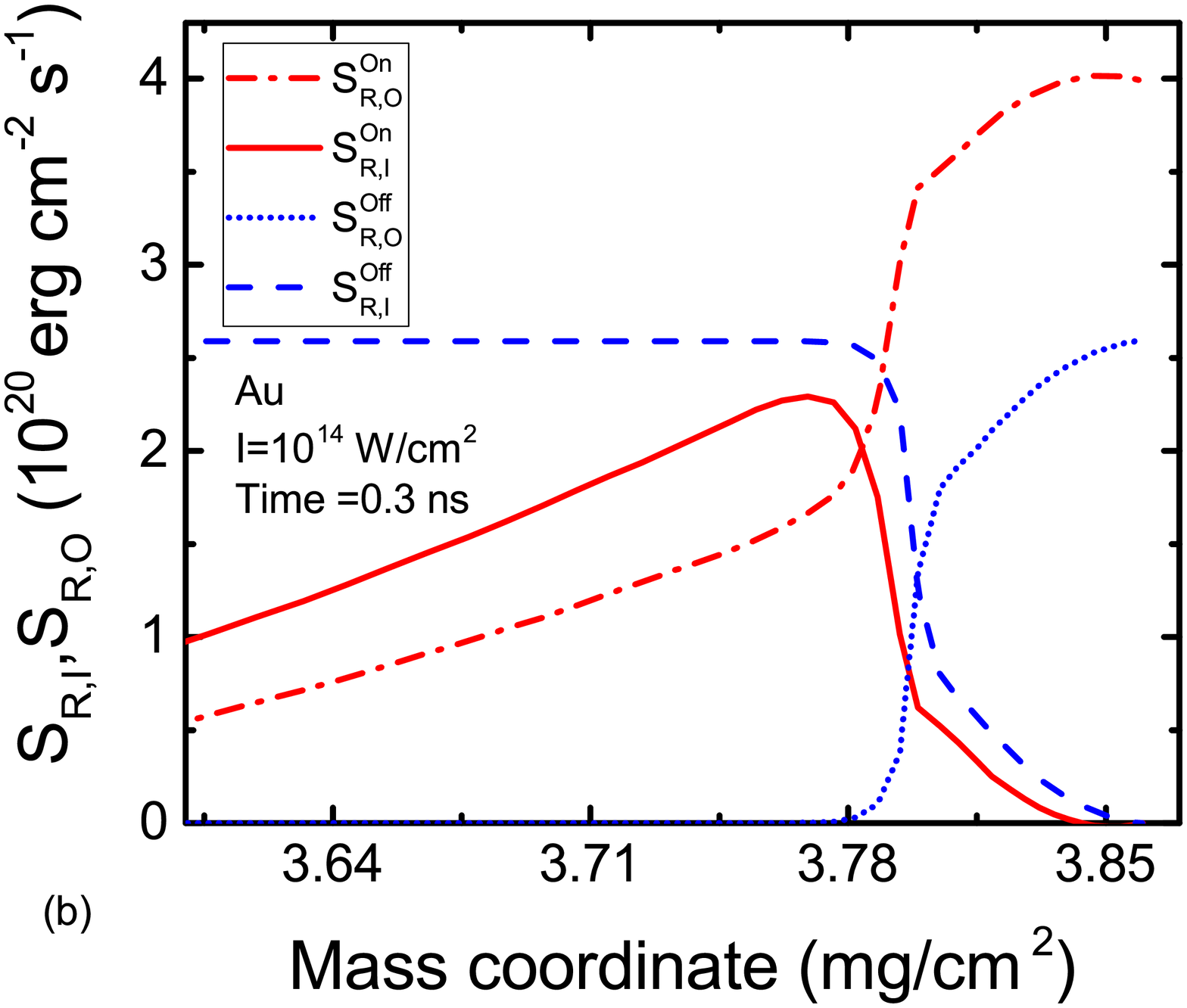}
\caption{(Colour online only) (a) Temperature variation in Au foil at 0.3 ns for laser intensity $10^{14}$ W/cm\tsu{2}. The solid red and dashed blue lines correspond to situations when absorption of radiation is switched on and off. (b) Variation of outward and inward radiation fluxes close to conversion layer against mass coordinate. The red and blue set of lines correspond to situations when absorption of radiation is switched on and off.\label{AuAbsOnOff}}
\end{figure}

\begin{figure}
        \includegraphics[width=0.49\textwidth]{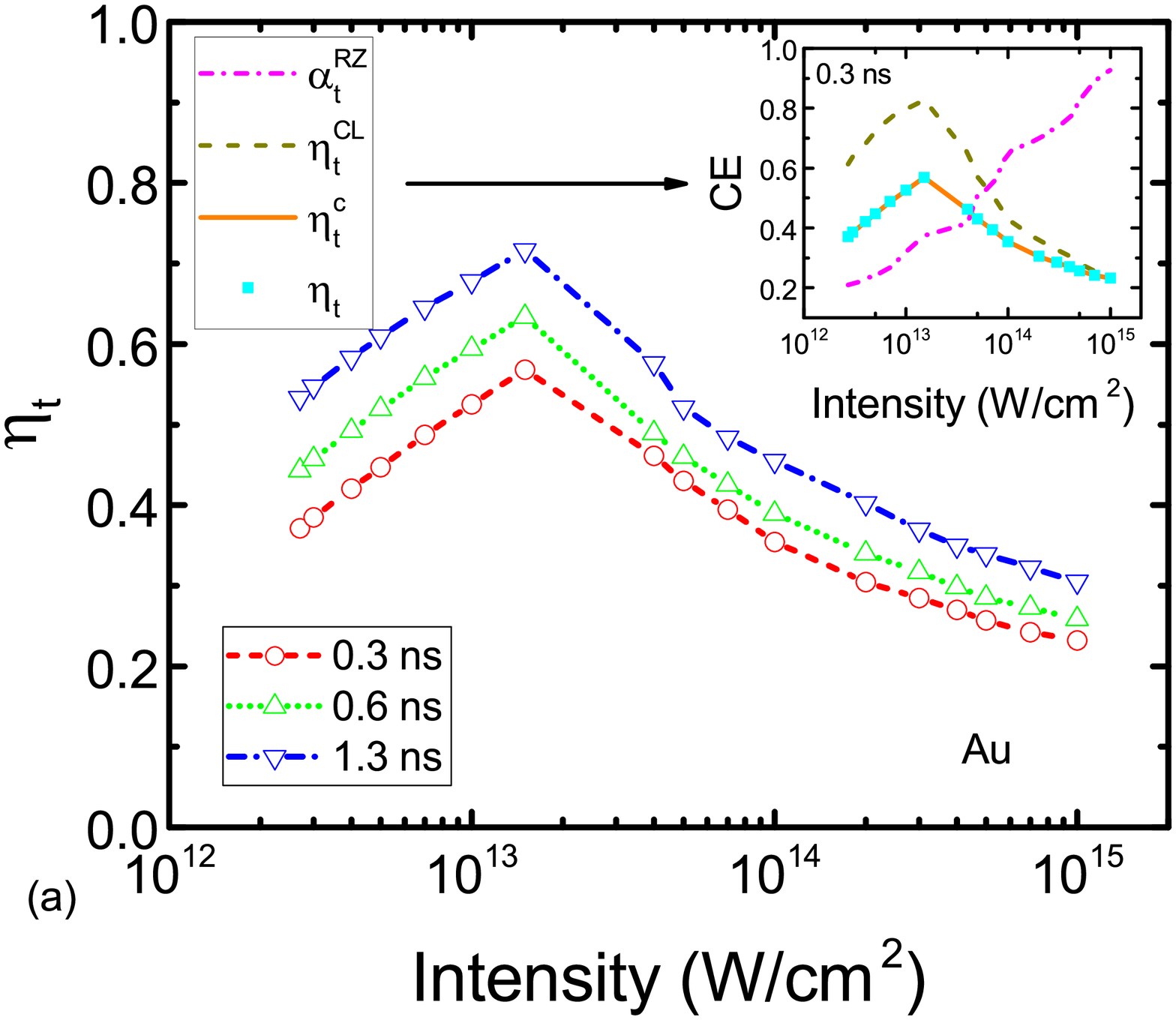}
        \includegraphics[width=0.49\textwidth]{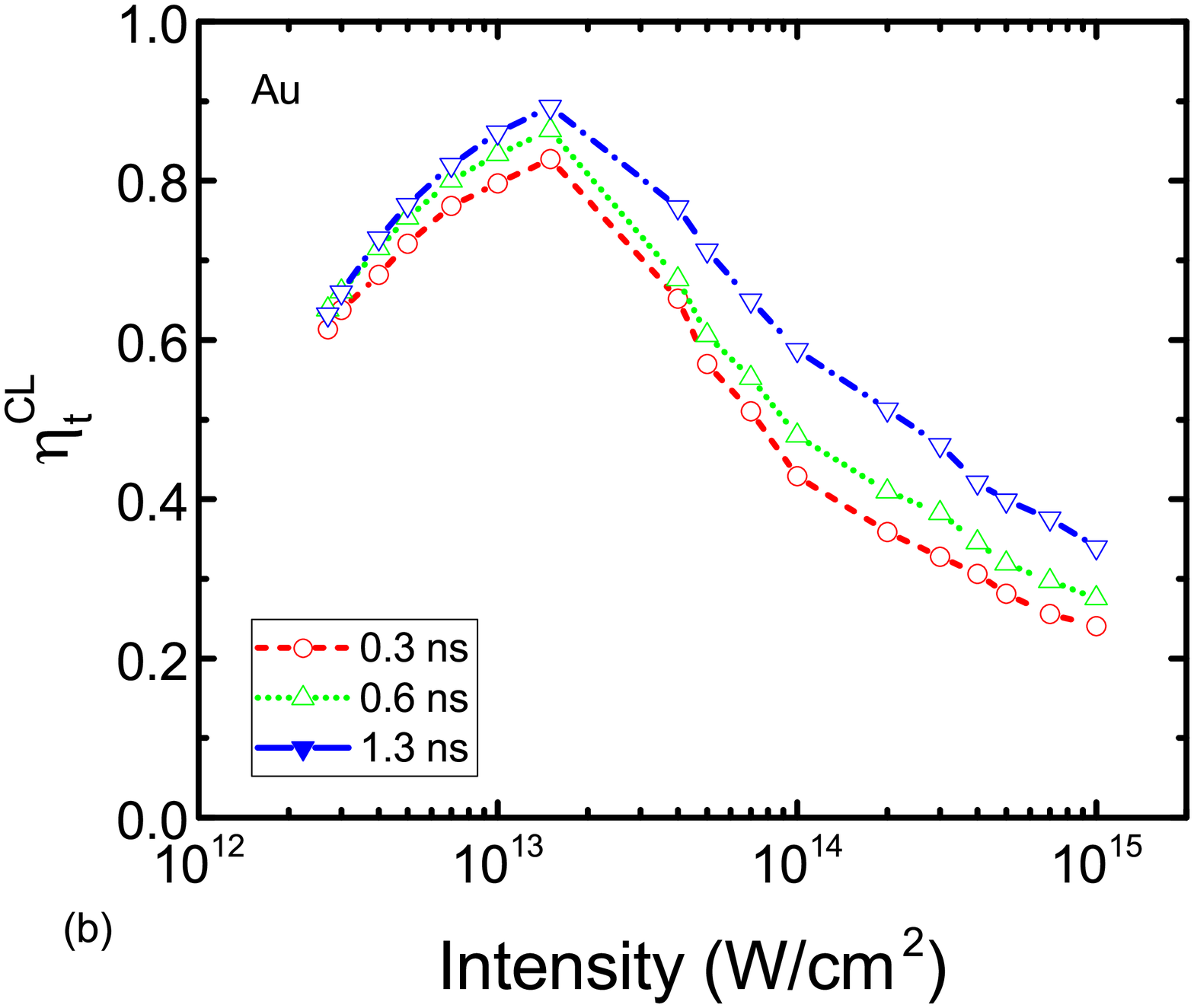}
        \includegraphics[width=0.49\textwidth]{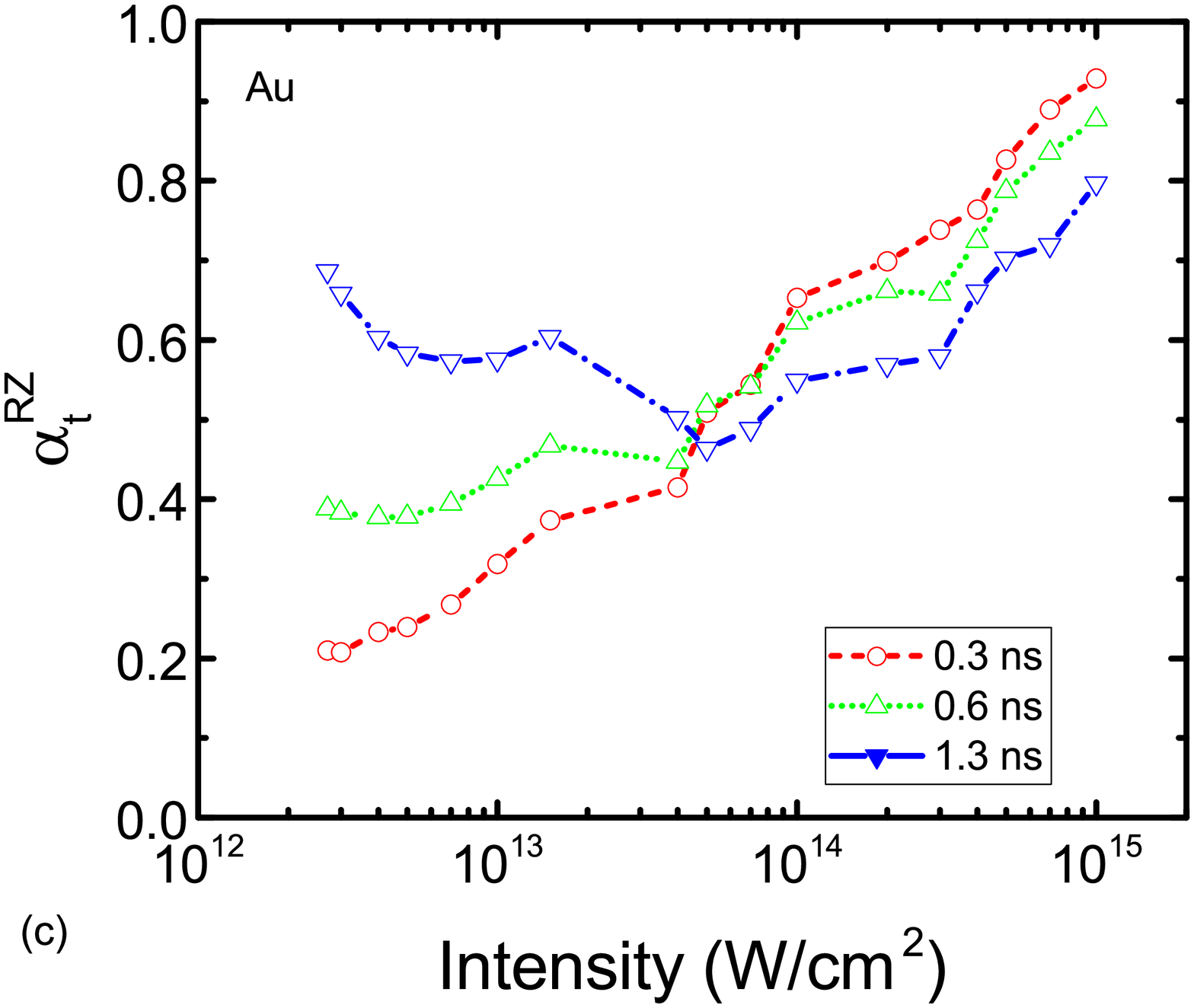}
\caption{(Colour online only) (a) Variation of conversion efficiency of Au with laser intensity at time instants of 0.3, 0.6 and 1.3 ns. In inset plot, emission characteristics of conversion layer and reemision zone along with $\eta$ simulated and calculated from Eq. \ref{EqEtat} are shown against laser intensity. Variation of (b) $\eta^{CL}_t$ and (c) $\alpha^{RZ}_t$ with intensity at 0.3 ns, 0.6 ns and 1.3 ns.\label{AuEtaAna}}
\end{figure}

\begin{figure}[]
        \includegraphics[width=0.49\textwidth]{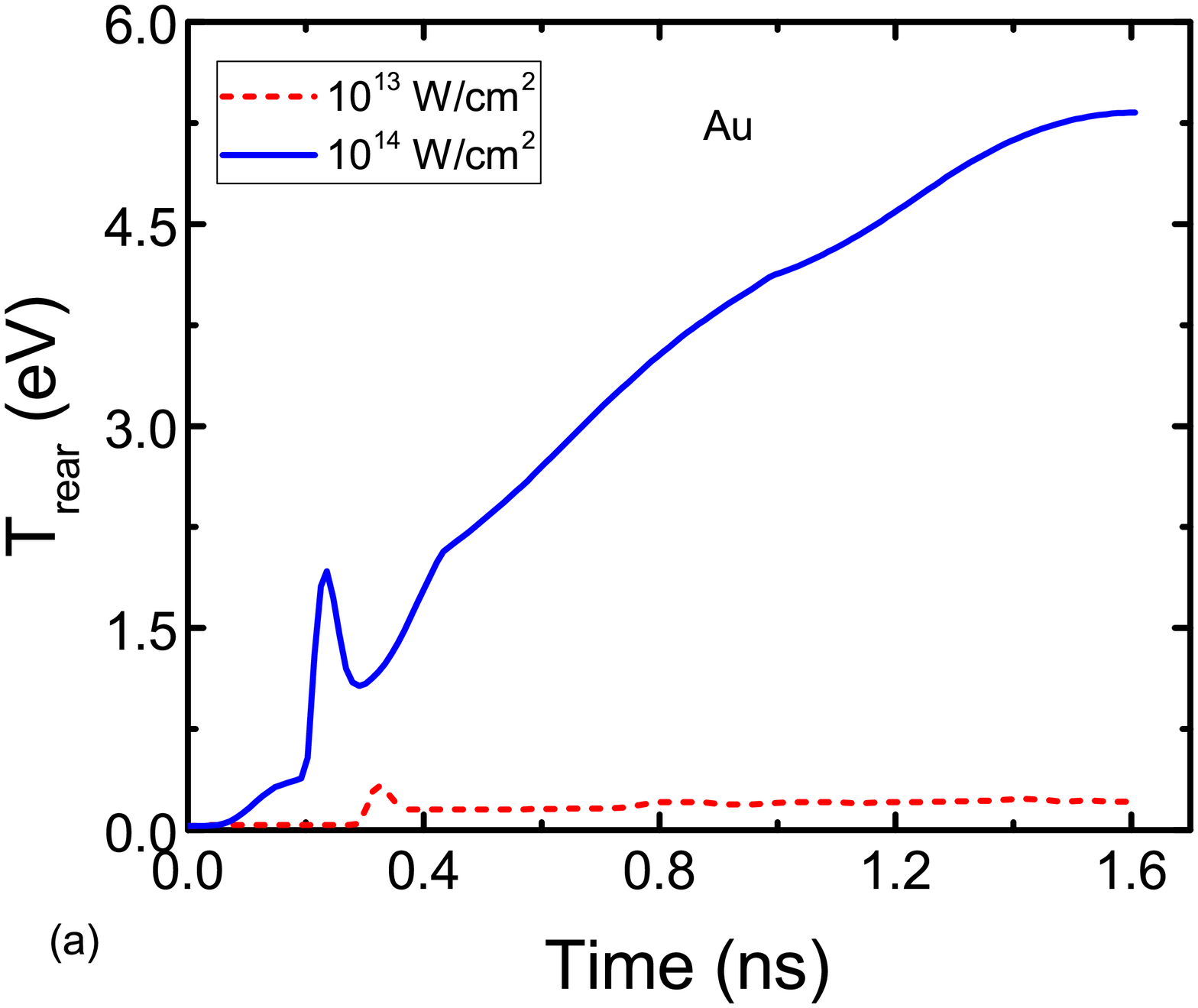}
        \includegraphics[width=0.49\textwidth]{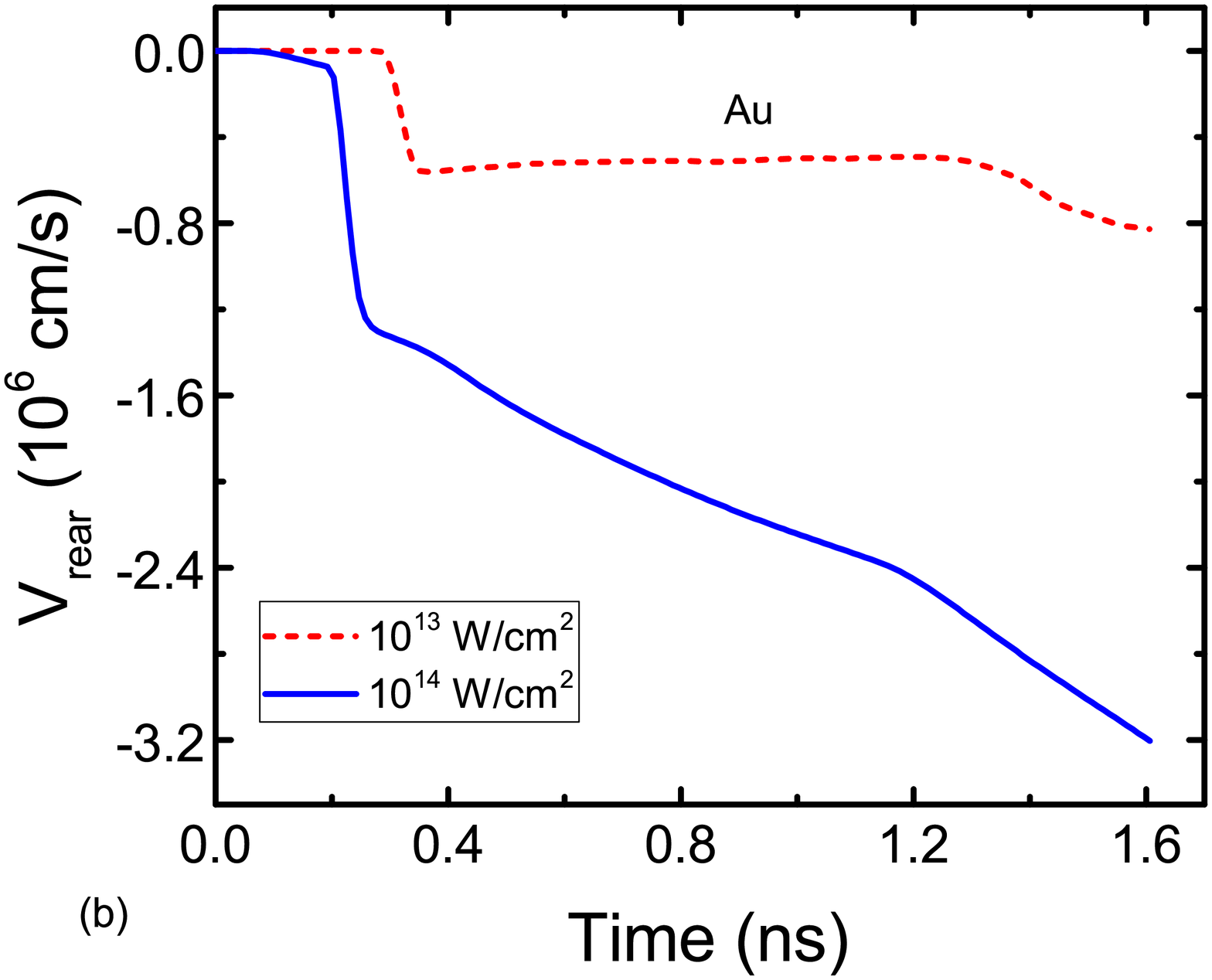}
\caption{(Colour online only) Rear surface (a) temperature and (b) velocity (in the unit of $10^{6}$ cm/s) as a function of time for Au. The two different sets of lines correspond to laser intensities of $10^{13}$ W/cm\tsu{2} (red colour) and $10^{14}$ W/cm\tsu{2} (blue colour).\label{AuTRear}}
\end{figure}

In this subsection, we will discuss in detail the characteristic behaviour of conversion efficiency with laser intensity for gold.  The total conversion efficiency of the foil is the sum of emission characteristics from CL and RZ \cite{ sigel1990conversion,eidmann1990conversion}. To determine the emission contributions from both regions, one has to numerically separate CL and RZ \cite{eidmann1990conversion}. This decomposition has been performed in our RHD simulations by artificially switching off the absorption of x-rays in the radiation and material energy equations while keeping the emission terms unaltered. The results of one such calculation is shown in Fig. \ref{AuAbsOnOff} (a).
Here, we have plotted electron temperature against mass coordinate at 0.3 ns in gold foil driven by laser of intensity $10^{14}$ W/cm\tsu{2}. It is observed that artificially switching off the absorption leads to disappearance of RZ whereas CL remains intact. In Fig. \ref{AuAbsOnOff} (b), we have shown the spatial variation of inward and outward fluxes in the two cases when absorption is on and off. With absorption off, both inward ($S^{Off}_{R,I}$) and outward fluxes ($S^{Off}_{R,O}$) approach same values on the two sides of the CL. This value of flux acts as source flux to drive radiation heat wave (RHW) in RZ when absorption is switched on. In that condition, outward flux ($S^{On}_{R,O}$) increases significantly in CL whereas inward flux ($S^{On}_{R,I}$) decreases sharply as we move away from CL. The results thus obtained are further used to determine the emission characteristics of both regions separately. In our simulations, since the conversion efficiencies  are defined as integrated quantities in energy and time, further analysis is performed with radiated energies instead of fluxes. In the CL, with laser acting as an input source, equal emission of radiation is obtained in both inward and outward direction. Let us define conversion efficiency of CL till time t as $\eta^{CL}_t=\left(2E^{Off}_{R,O}/E_L\right)$ where $E_L$ is the laser energy and $E^{Off}_{R,O}$ is the outward radiated energy when absorption is off. In reemission zone, inwards radiated energy $E^{Off}_{R,I}$ acting as input source leads to formation of RHW. To analyze the emission characteristics of RZ, we define the average albedo till time t as $\alpha^{RZ}_t=\left[\left(E^{On}_{R,O}/E^{Off}_{R,O}\right)-1\right]$, where $E^{On}_{R,O}$ is the outward energy when absorption is on. The total conversion efficiency ($\eta{^c_t}$) integrated till time t can be obtained from corresponding integrated quantities $\eta^{CL}_t$ and $\alpha^{RZ}_t$ as
\be
\eta{^c_t}=\frac{\eta^{CL}_t}{2}+\alpha^{RZ}_t\frac{\eta^{CL}_t}{2}=\frac{\eta^{CL}_t}{2}\left(1+\alpha^{RZ}_{t}\right).
\label{EqEtat}
\ee      
In Fig. \ref{AuEtaAna} (a), we have shown the variation of conversion efficiency ($\eta_t$ from simulations) against laser intensity at three different time instants 0.3, 0.6 and 1.3 ns. 
As time of observation increases, $\eta_t$ enhances at all intensities with maximum lying at the same value of intensity. In the inset plot, the variation of emission characteristics of CL and RZ ($\eta{^{CL}_t}$ and $\alpha{^{RZ}_t}$) along with conversion efficiency ($\eta_t$ from simulations and $\eta{^c_t}$ calculated from Eq.\ref{EqEtat}) with intensity are shown  at a representative time of 0.3 ns. We observe that numerical separation of radiating plasma into two different regions is well justified as the results from simulations (scattered points) and from Eq.\ref{EqEtat} (solid line) are found to be the same. Similar kind of results were also obtained for instantaneous conversion efficiencies from RHD simulations when evaluated in terms of fluxes instead of integrated energy values \cite{eidmann1990conversion}. To understand the behaviour of conversion efficiency in terms of decomposed CL and RZ, we have shown the variation of $\eta^{CL}_{t}$ and $\alpha^{RZ}_{t}$ with intensity in Figs. \ref{AuEtaAna} (b) and (c) at time instants of 0.3, 0.6 and 1.3 ns. We note that $\eta^{CL}_{t}$ shows the same trend with intensity as depicted by $\eta_t$. For lower intensities, conditions in CL satisfy LTE conditions and emission only depends upon the temperature. So $\eta^{CL}_{t}$ increases with laser intensity. After reaching a maximum, the conditions in CL make a transition from LTE to NLTE. For higher intensities, more energy is transported into the foil due to convective motion of material. As a result, less energy is radiated towards laser side, thus leading to further reduction of $\eta^{CL}_{t}$ with intensity. On the other hand, $\alpha^{RZ}_{t}$ shows opposite behaviour with time on either side of a transition intensity $\sim 4\times 10^{13}$ W/cm\tsu{2} as shown in Fig. \ref{AuEtaAna} (c). For intensity values lower than transition value, $\alpha^{RZ}_{t}$ increases with time but the gap reduces for intensities approaching the transition intensity. After the transition, the trend reverses and $\alpha^{RZ}_t$ decreases with time as we move beyond transition intensity. At later times, RZ is not able to contain the M-band radiation emerging from CL at higher intensities, so $\alpha^{RZ}_t$ reduces with time. To further strengthen this argument, we have investigated the temporal variation of rear surface preheating due to M-band emission under different laser intensities. In Figs. \ref{AuTRear} (a) and (b), we have shown rear surface temperature ($T_{rear}$) and velocity ($V_{rear}$) variation with time for two different laser intensities ($10^{13}$ W/cm\tsu{2} and $10^{14}$ W/cm\tsu{2}).
For the lower value of incident laser intensity ($10^{13}$ W/cm\tsu{2}), preheating of the rear surface due to hard x-rays is negligible. But at the higher intensity value of $10^{14}$ W/cm\tsu{2}, the M-band emission from hot conversion layer reaches the rear surface leading to further increase in rear surface temperature and velocity. 

\subsubsection{Comparison of x-ray conversion efficiency among the elements}

\begin{figure}
        \includegraphics[width=0.49\textwidth]{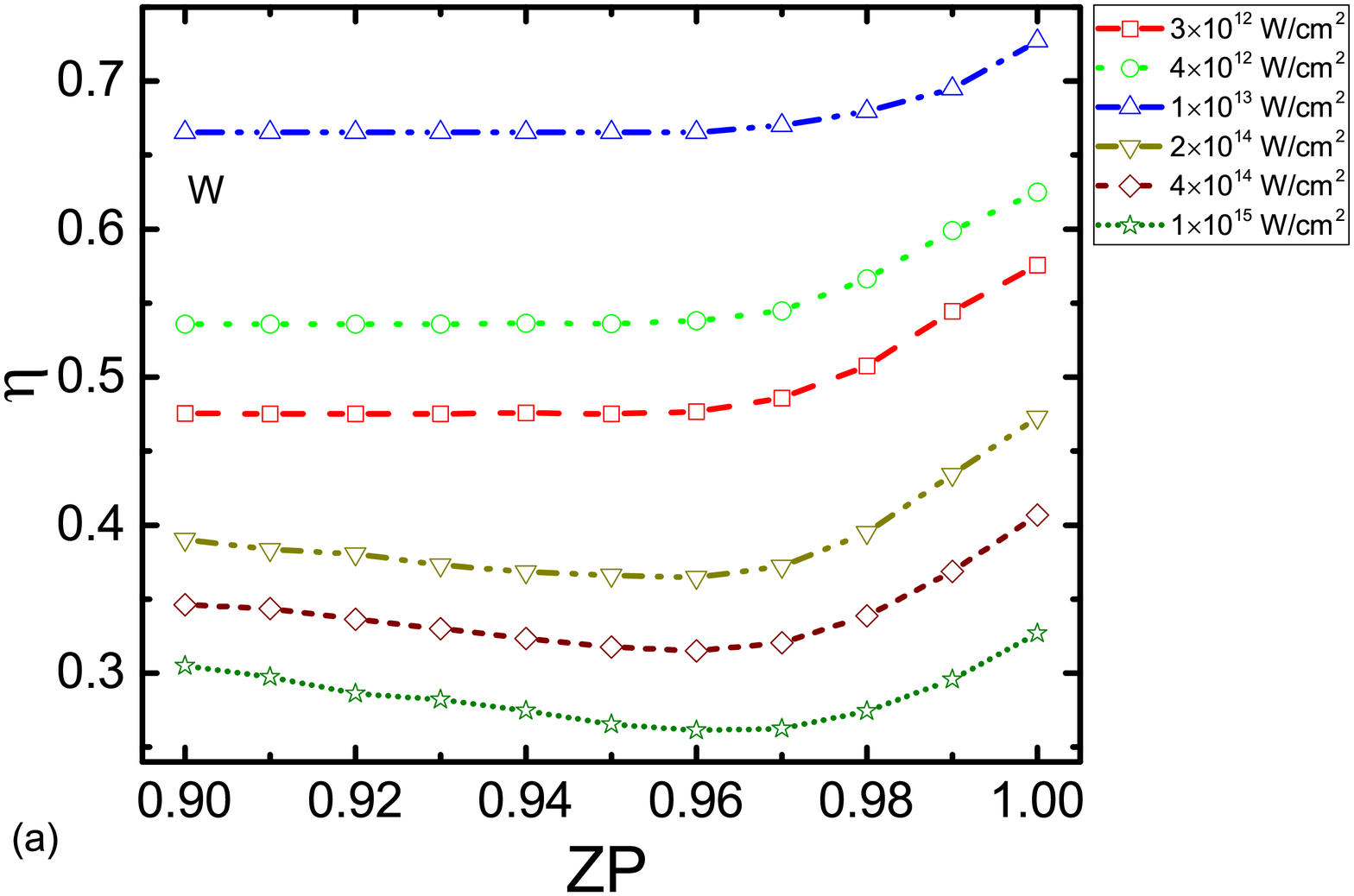}
        \includegraphics[width=0.49\textwidth]{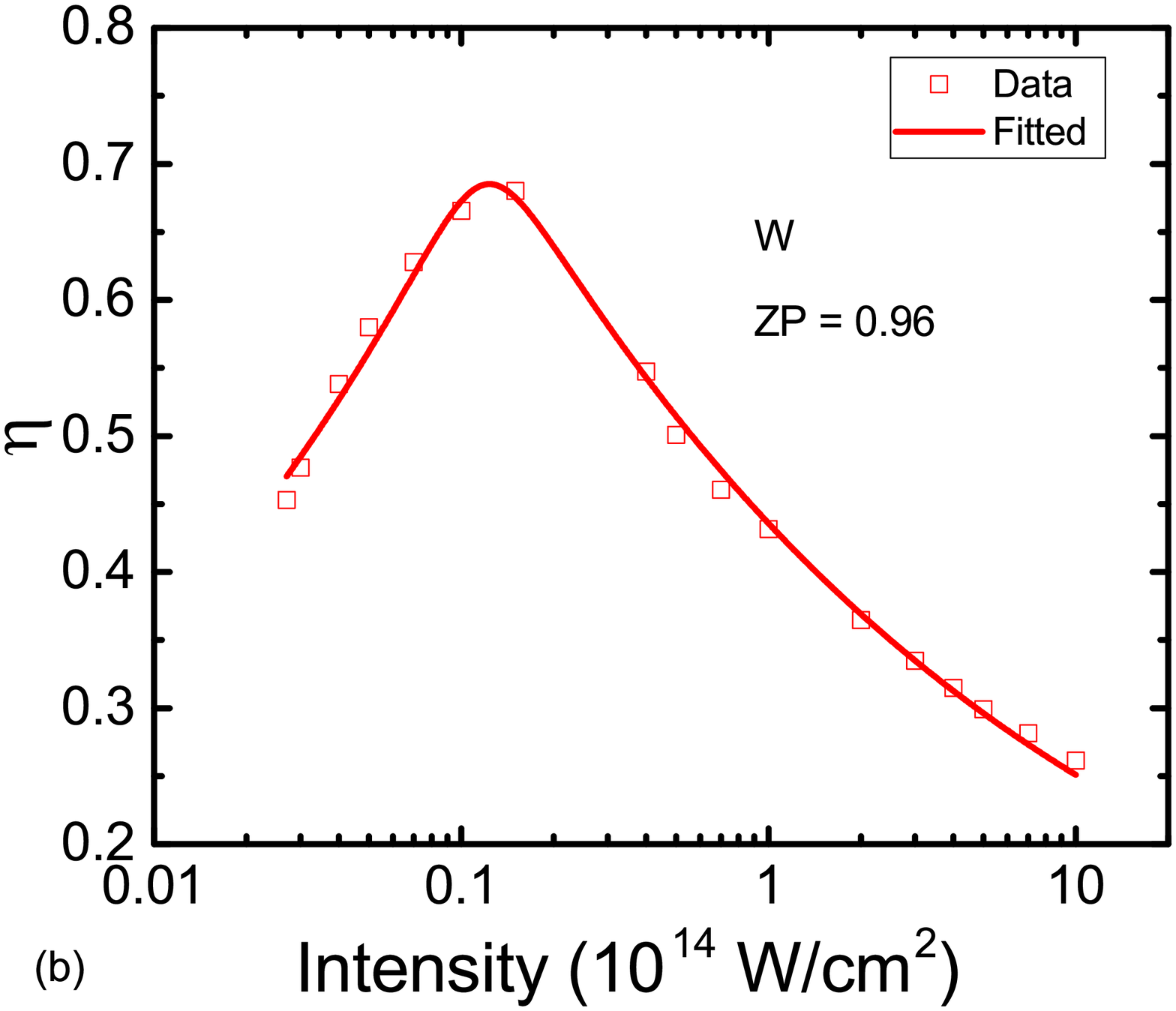}
        \includegraphics[width=0.49\textwidth]{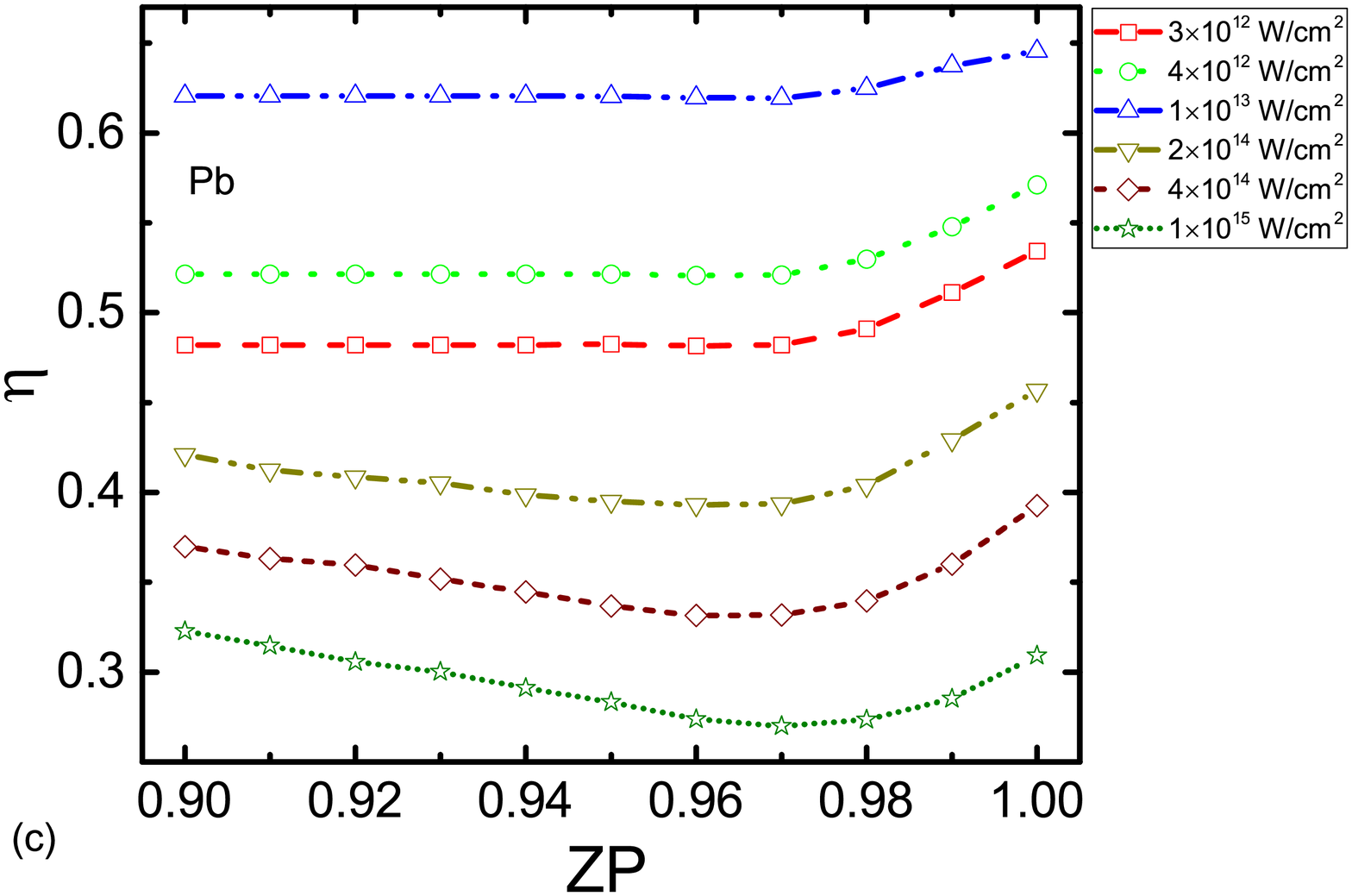}
        \includegraphics[width=0.49\textwidth]{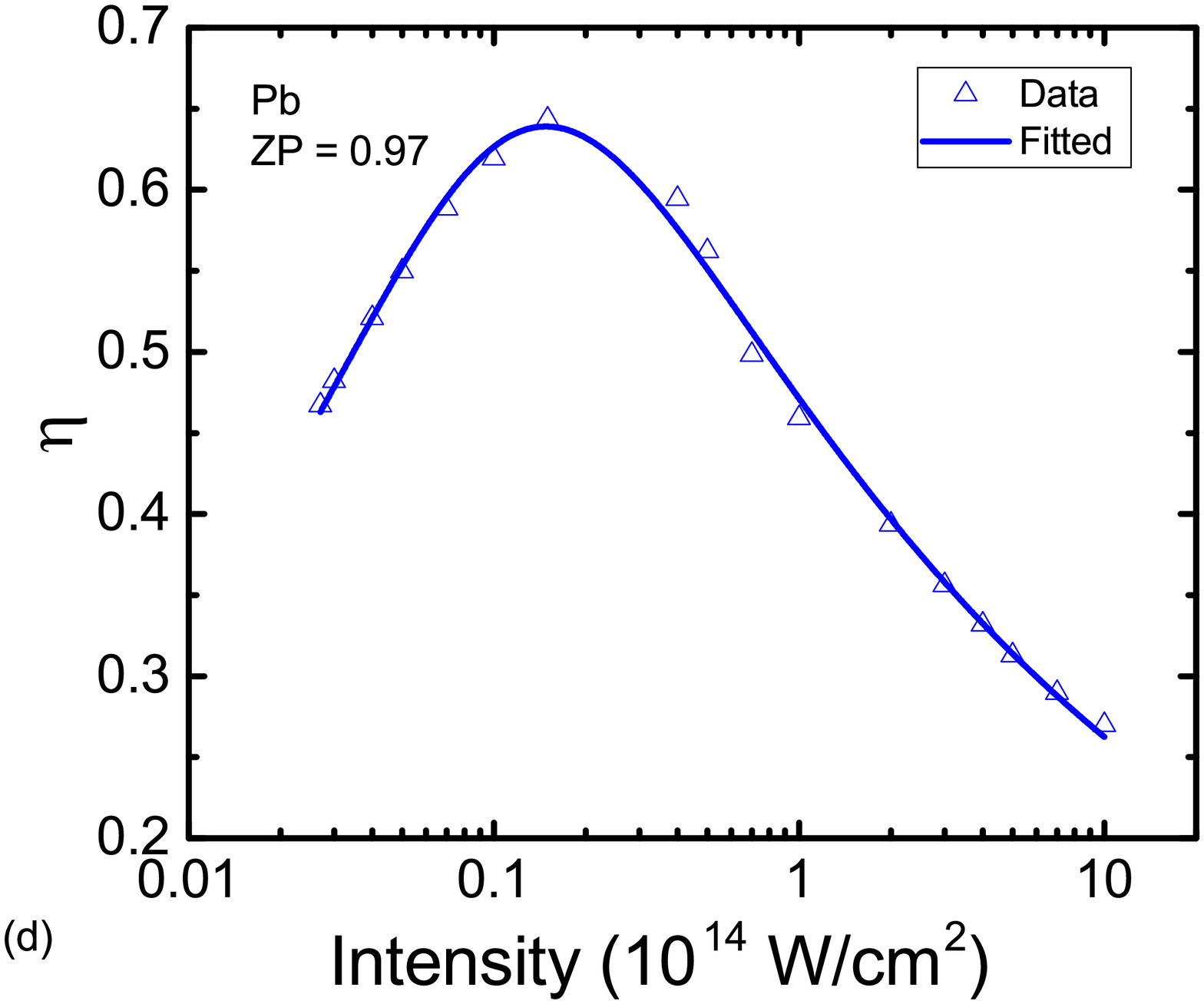}
        \includegraphics[width=0.49\textwidth]{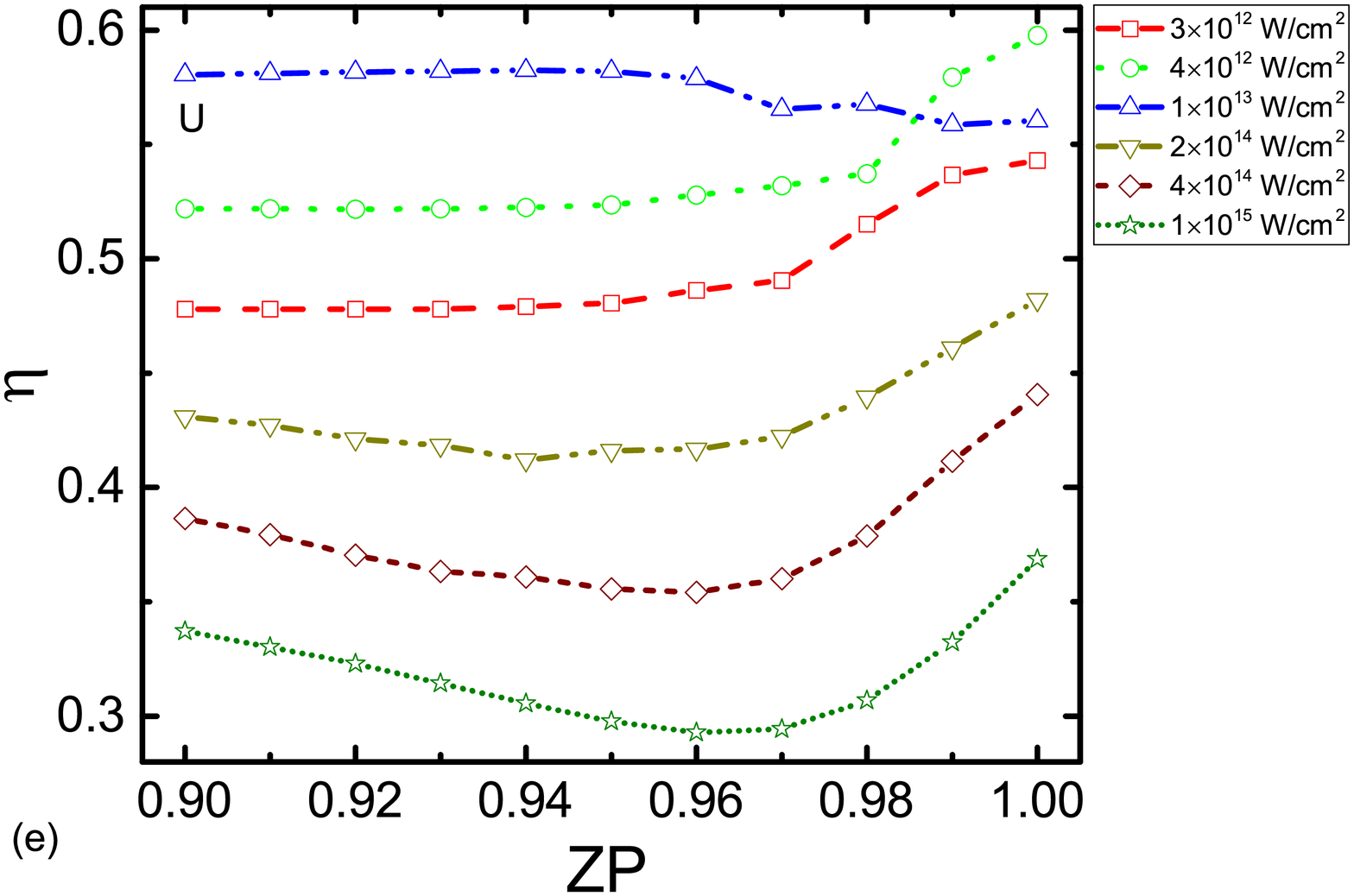}
        \includegraphics[width=0.49\textwidth]{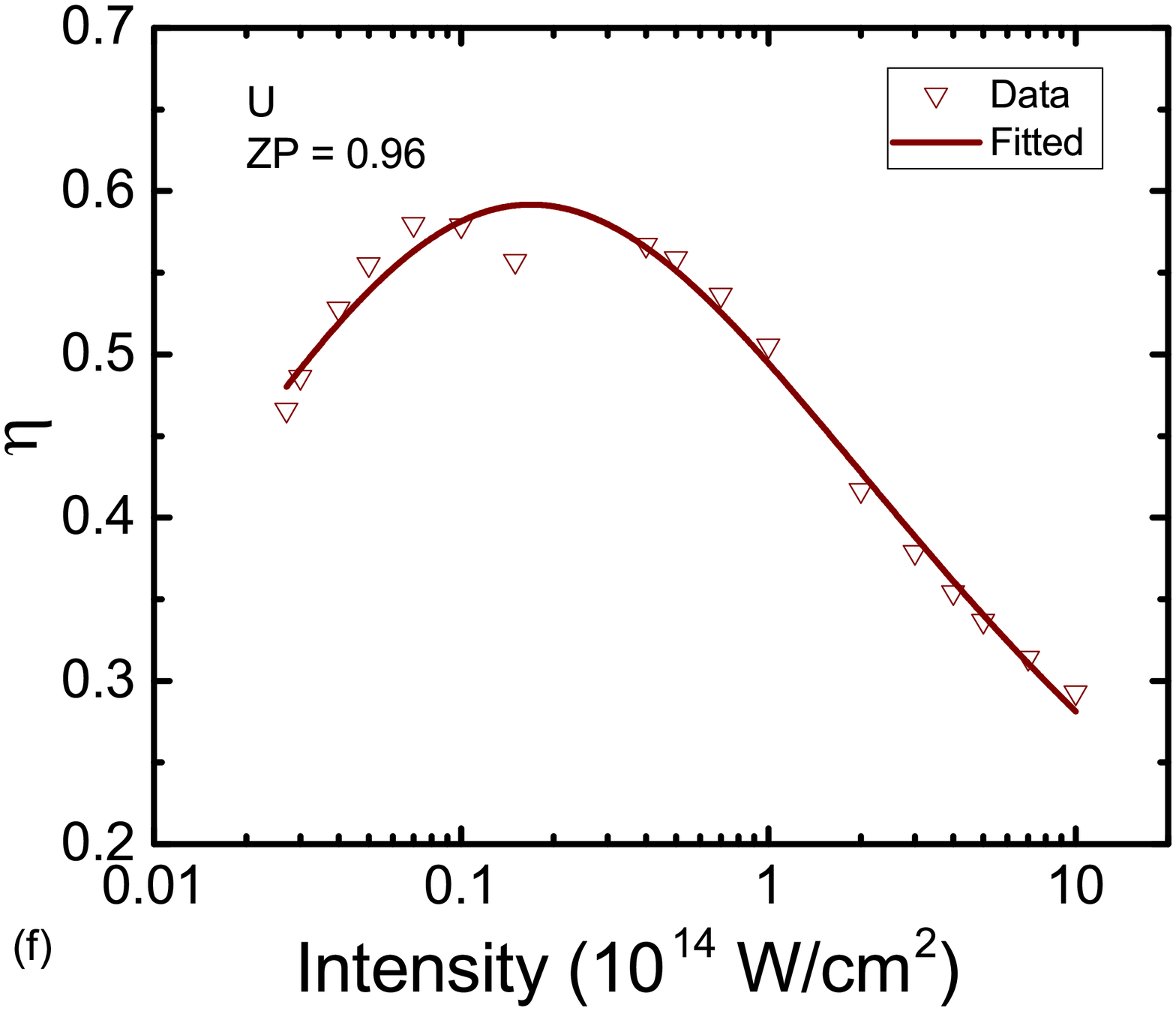}
\caption{(Colour online only) Conversion efficiency of (a) W, (c) Pb and (e) U foils driven by flat top 1 ns laser pulse as a function of zoning parameter at different laser intensities. The simulated $\eta$ data and fitted results from Eq.\ref{EqEtaFit} are shown for W, Pb and U in (b), (d) and (f) against laser intensity (in the unit of $10^{14}$ W/cm\tsu{2}). \label{AllEta}}
\end{figure}

\begin{figure}
        \includegraphics[width=0.49\textwidth]{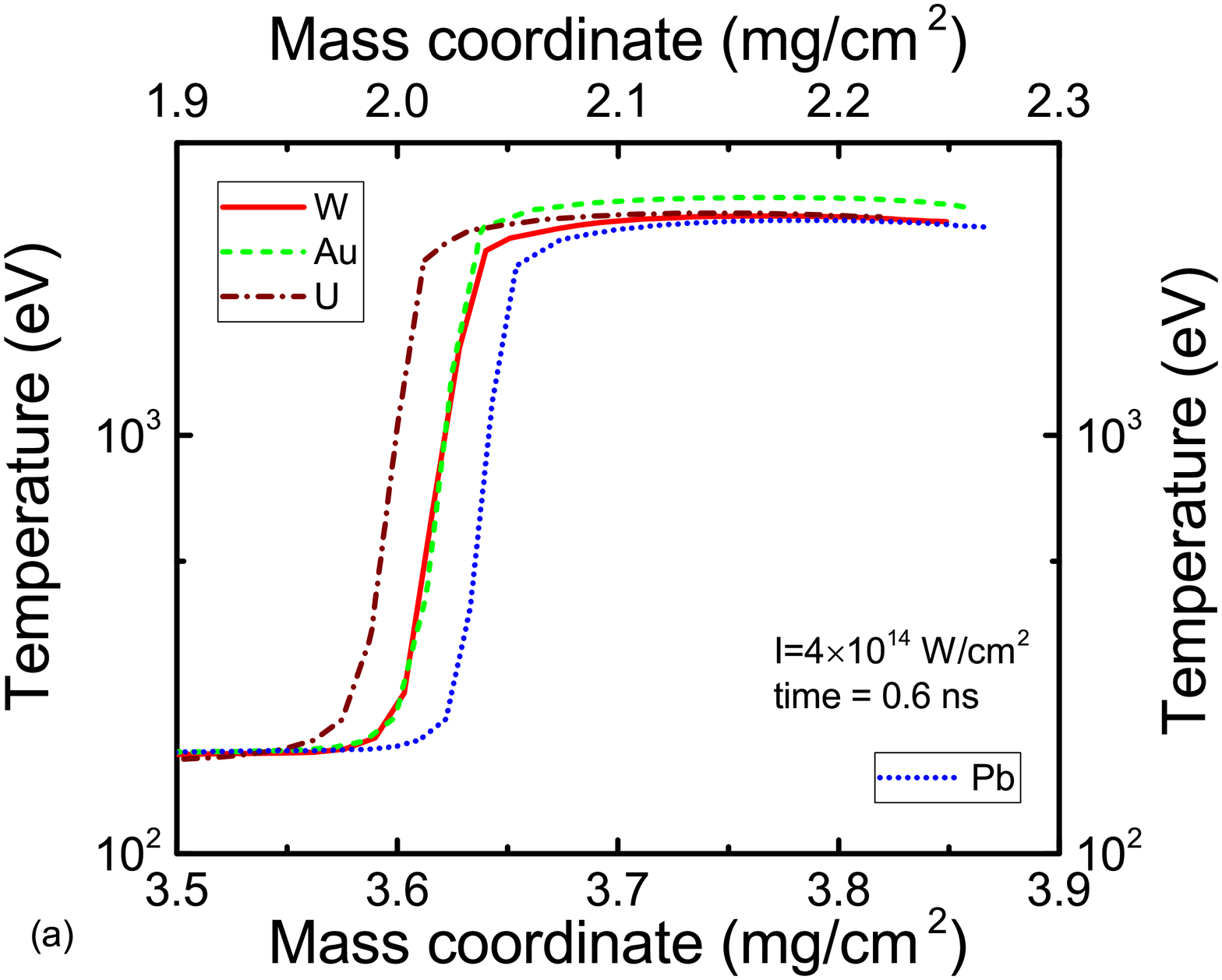}
        \includegraphics[width=0.49\textwidth]{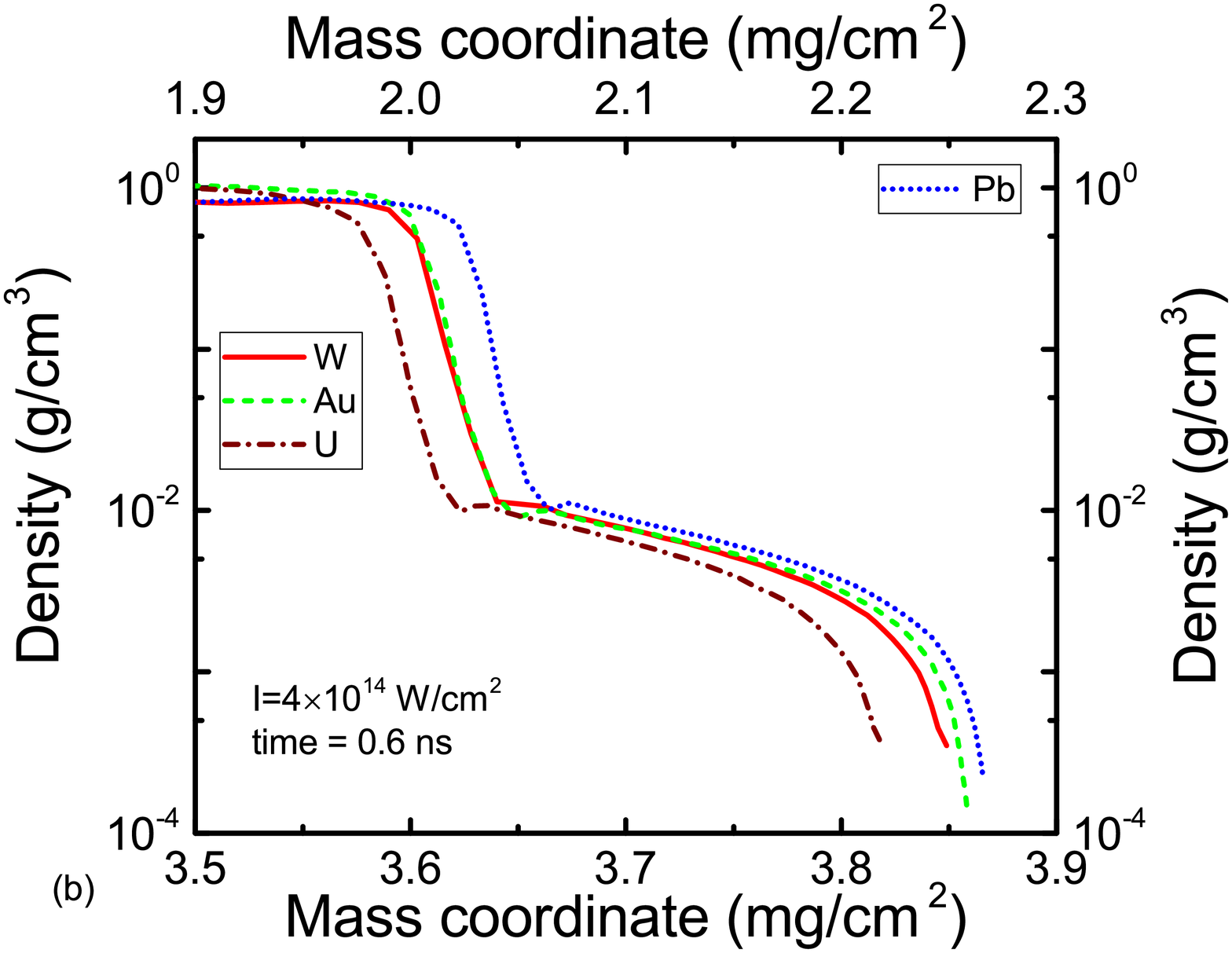}
\caption{(Colour online only) (a) Temperature and (b) density variation with mass coordinate in region close to conversion layer for all materials at 0.6 ns. All of the foils are irradiated with a representative laser intensity of $4\times 10^{14}$ W/cm\tsu{2}.\label{TempDensAll-0.6ns}}
\end{figure}

\begin{figure}
        \includegraphics[width=0.49\textwidth]{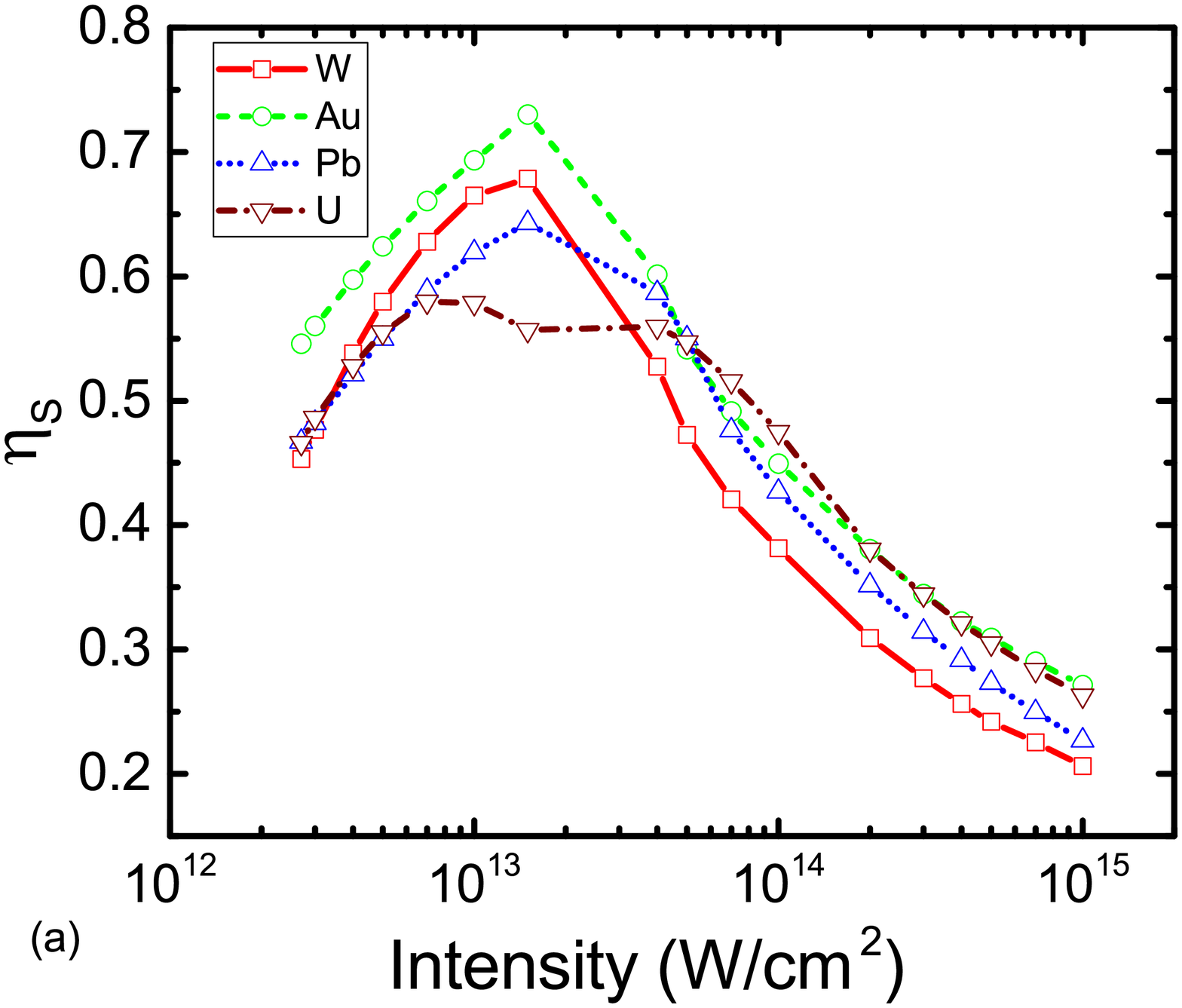}
        \includegraphics[width=0.49\textwidth]{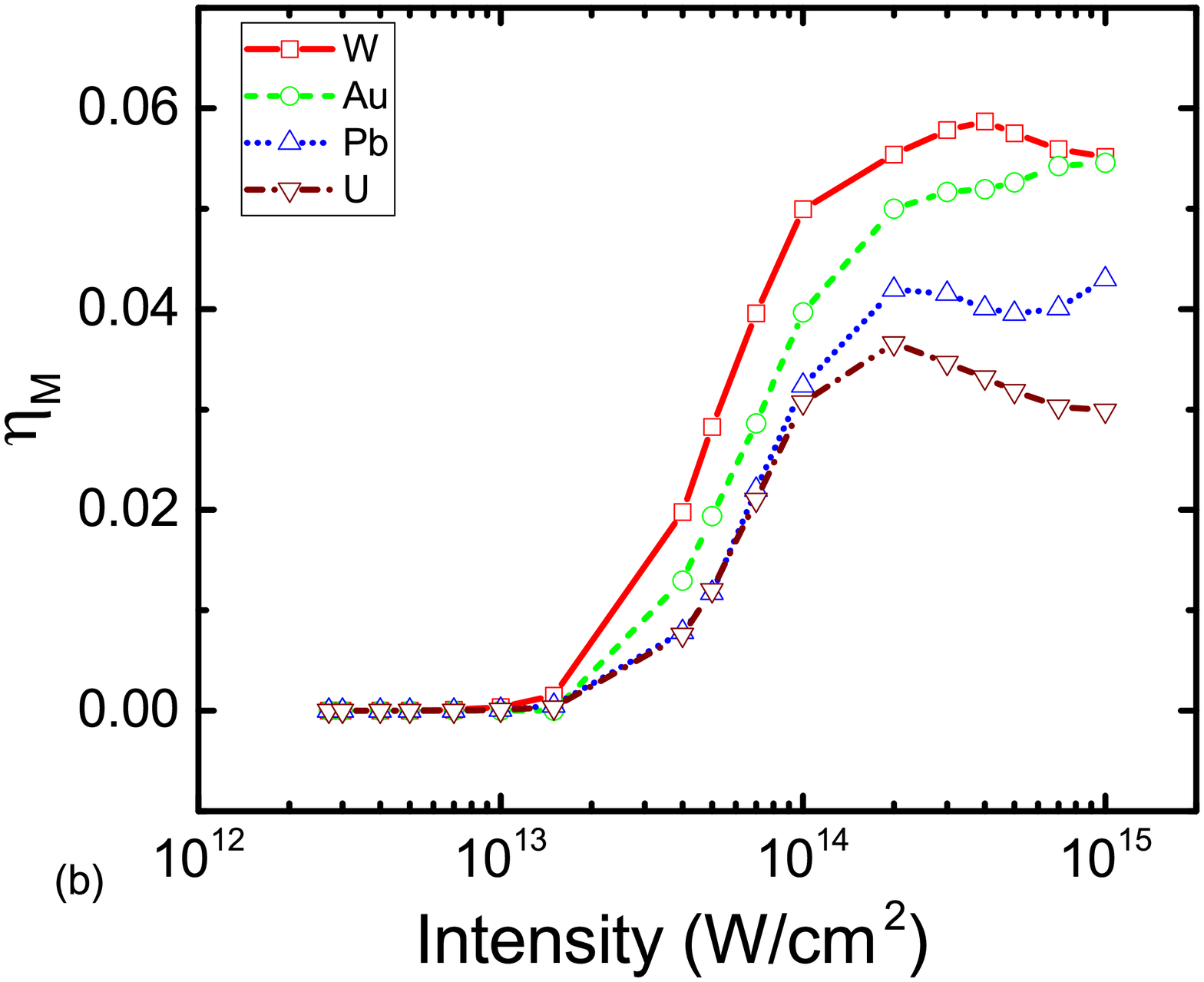}
\caption{(Colour online only) Conversion efficiency of different high-Z materials in photon energy range of (a) 0.15 keV to 1.6 keV and (b) 1.6 keV to 5 keV as a function of laser intensity.\label{EtaSMall}}
\end{figure}

\begin{figure}
        \includegraphics[width=0.49\linewidth]{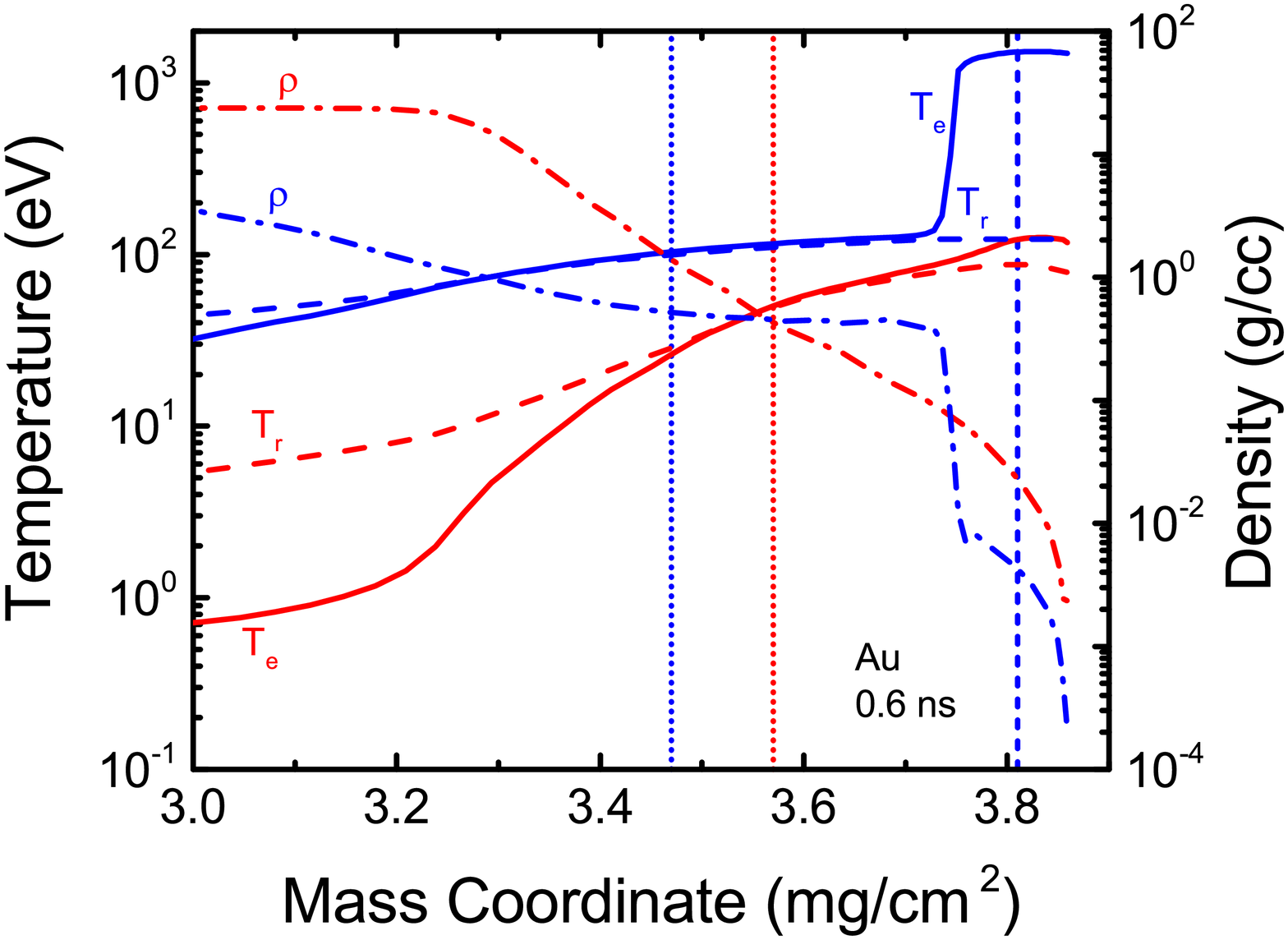}
\caption{(Colour online only) Electron temperature ($T_e$), radiation temperature ($T_r$) and density ($\rho$) variation against mass coordinate for laser irradiated gold foil at 0.6 ns. The vertical lines show the representative locations of conversion layer and reemission zone of the laser plasma region. Red and blue colour correspond to laser intensities of $10^{13}$ and $10^{14}$ W/cm\tsu{2}, respectively.}
\label{TRwithMassI}
\end{figure}

\begin{figure}
     \includegraphics[width=0.49\textwidth]{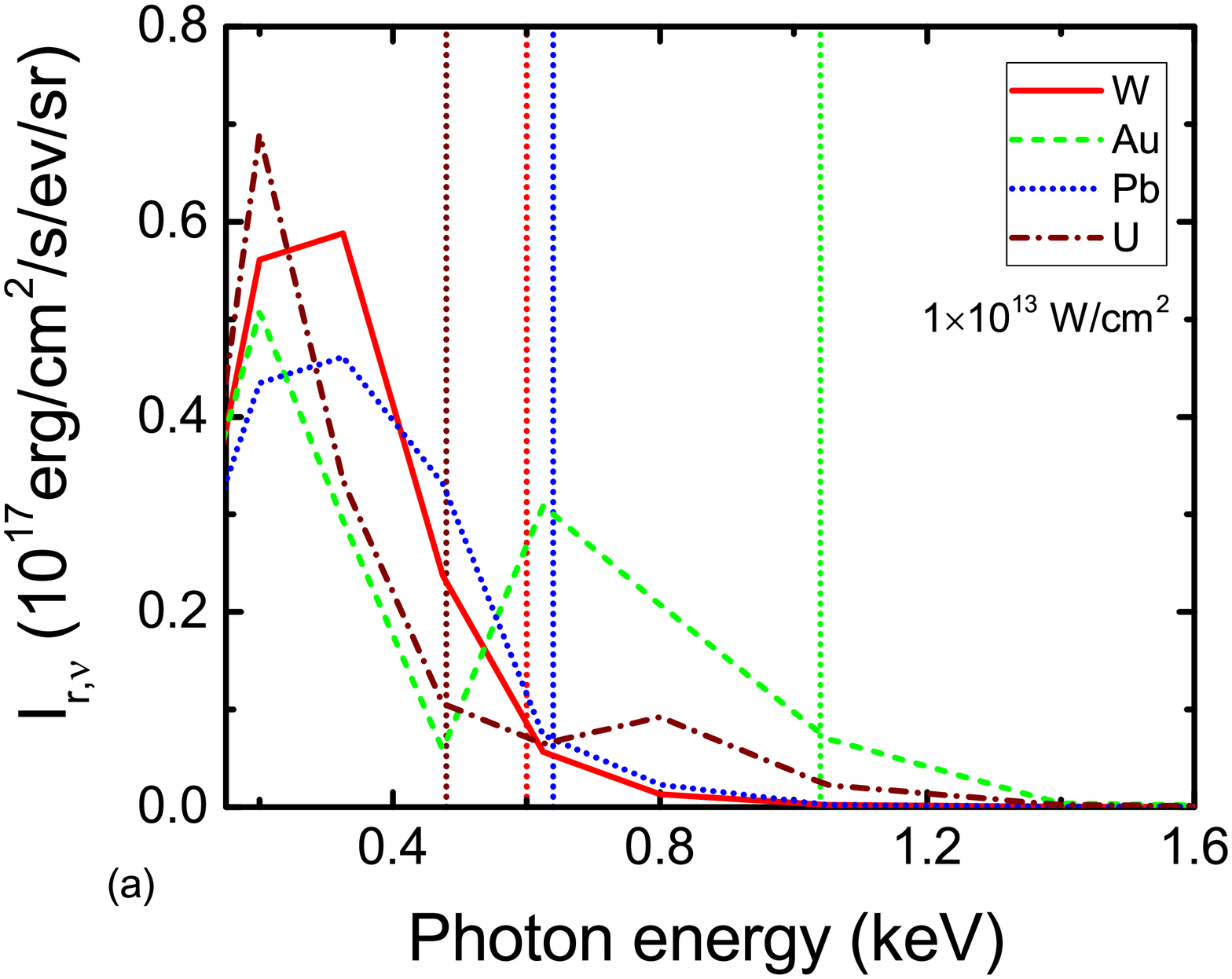}        \includegraphics[width=0.49\textwidth]{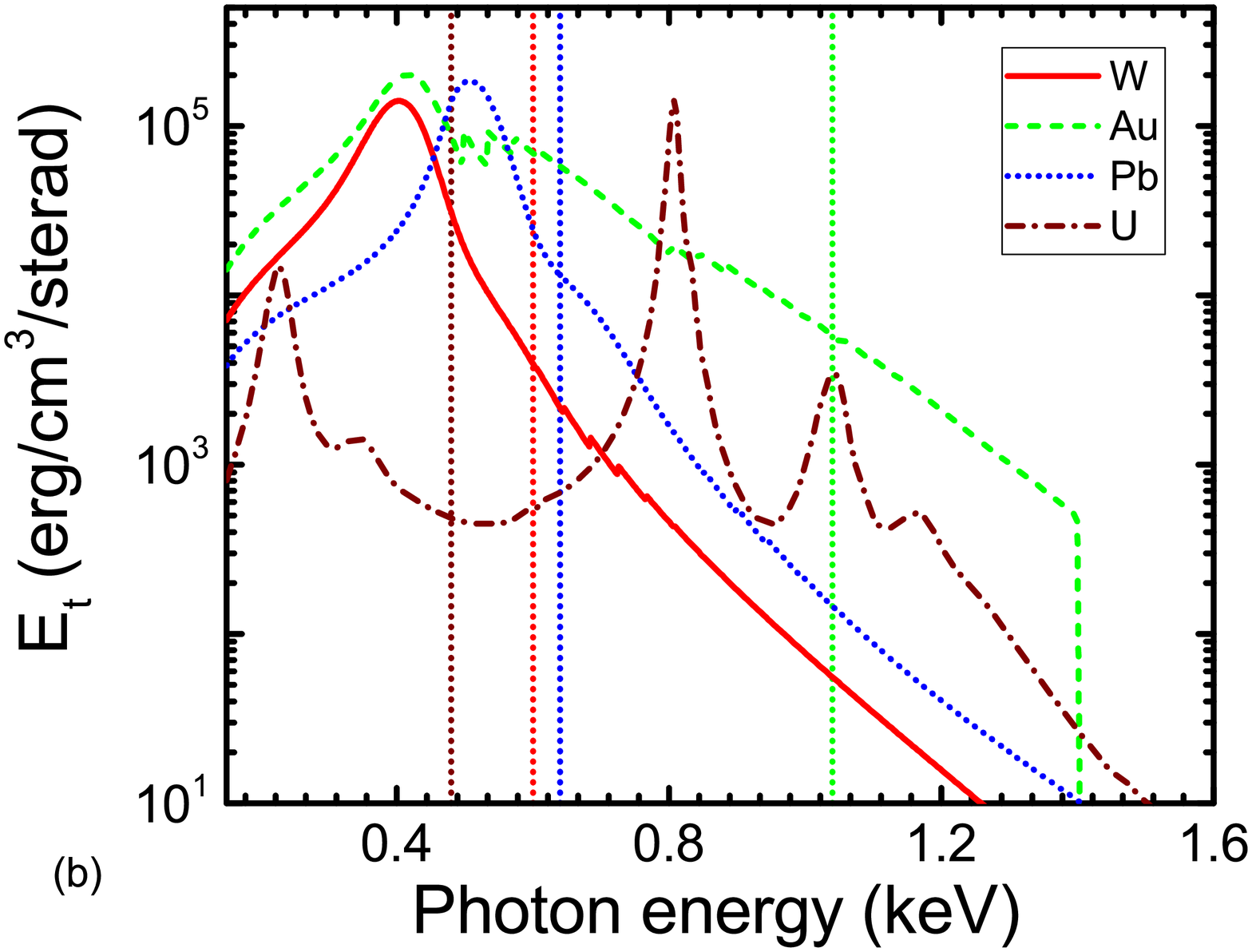}        \includegraphics[width=0.49\textwidth]{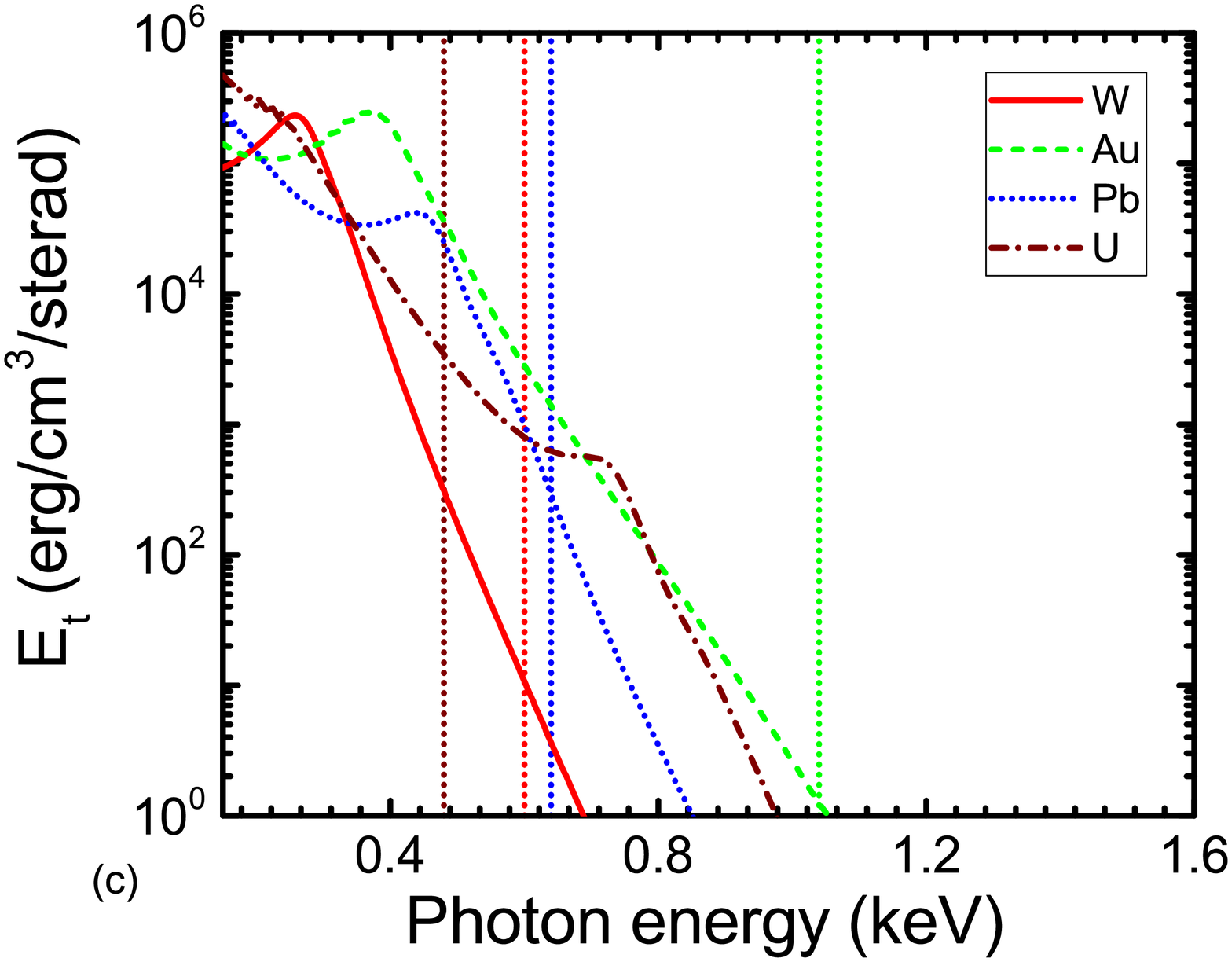}
\caption{(Colour online only) (a) Radiation spectrum and emission coefficients evaluated at ($\rho$,$T_{e}$) of (b) conversion layer and (c) reemission zone against photon energy range 0.15-1.6 keV for different high-Z materials irradiated by laser intensity of $10^{13}$ W/cm\tsu{2}. The different vertical lines represent photon cut-off energies.\label{FS1e13-1}}
\end{figure}

\begin{figure}
        \includegraphics[width=0.49\textwidth]{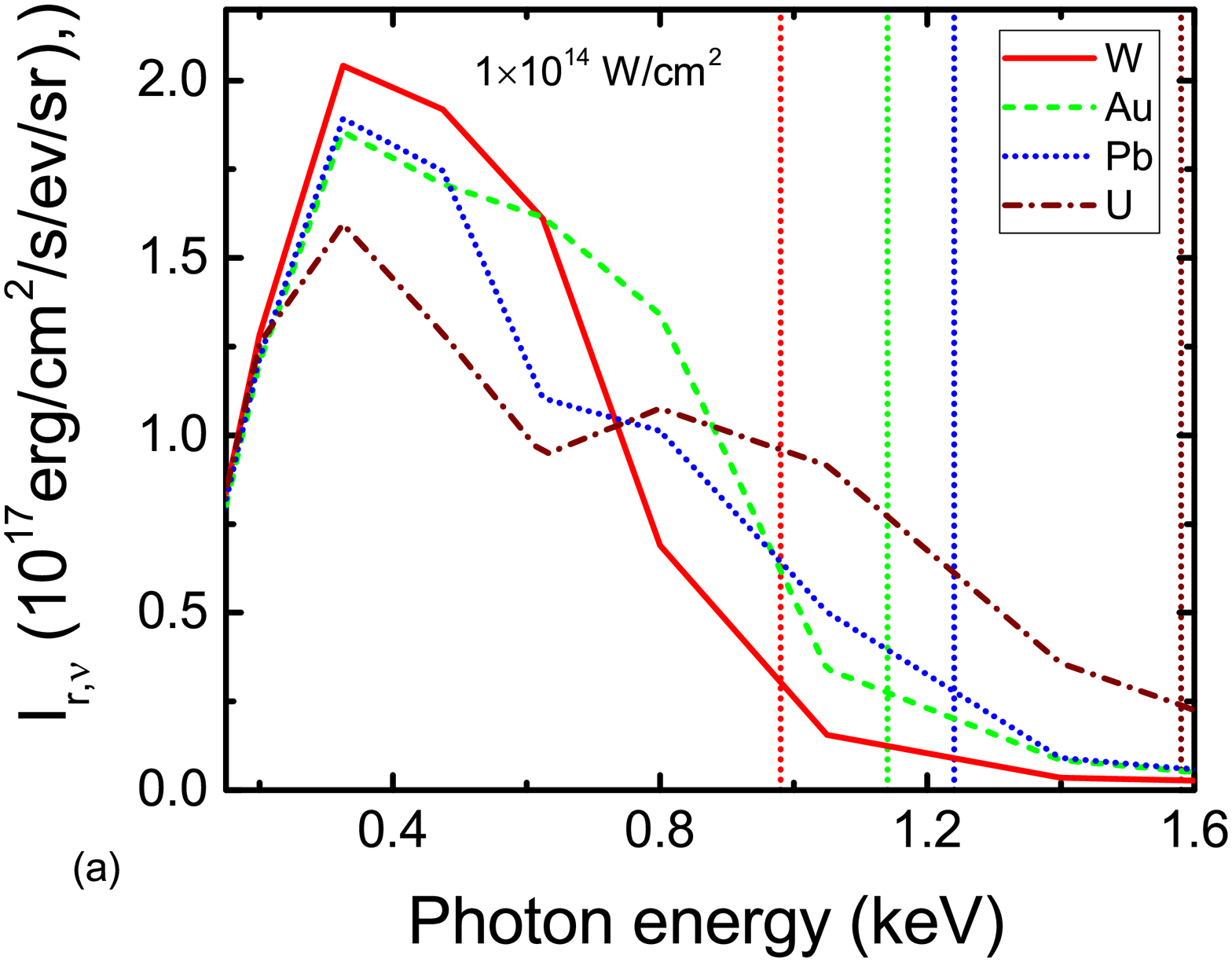}
        \includegraphics[width=0.49\textwidth]{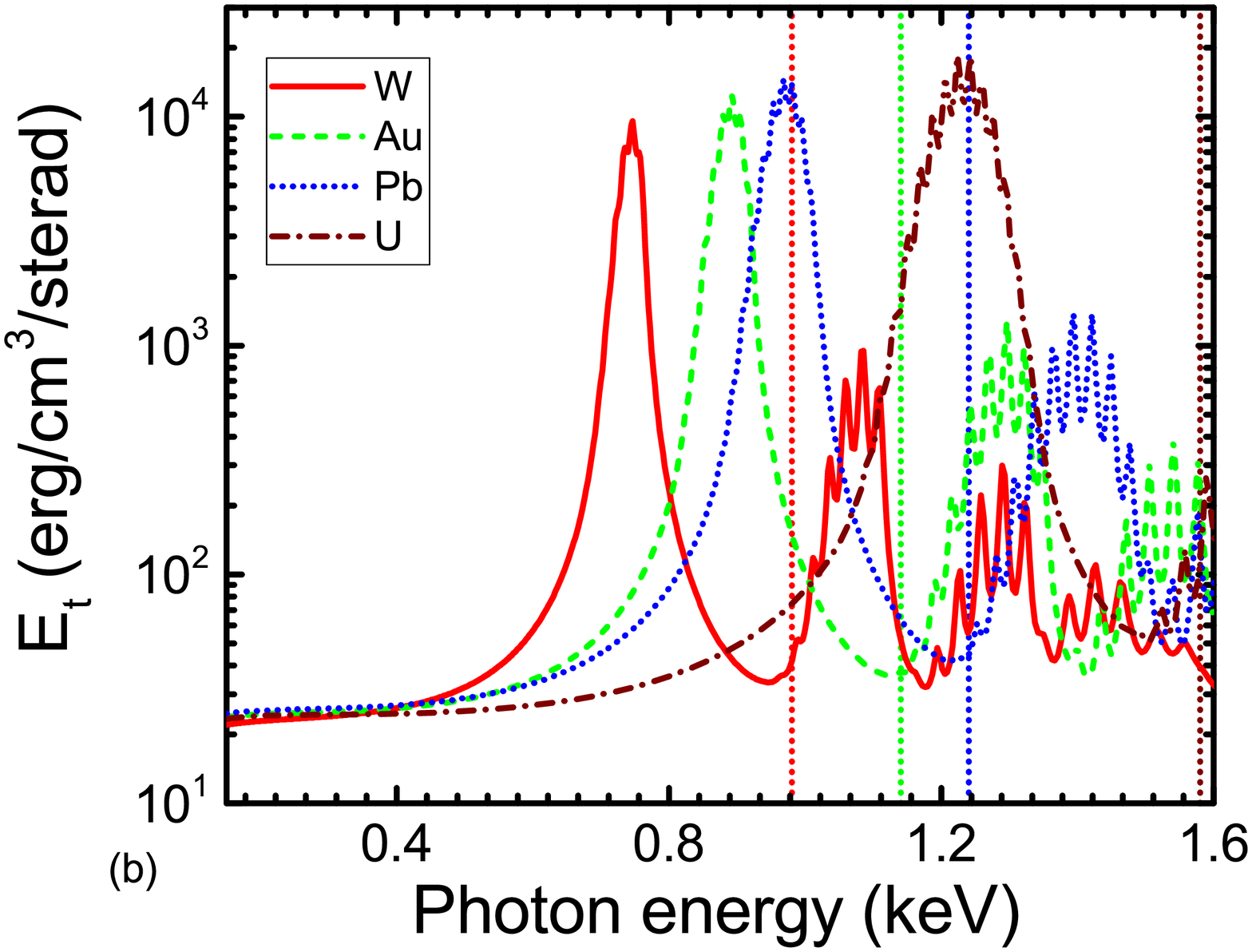}
        \includegraphics[width=0.49\textwidth]{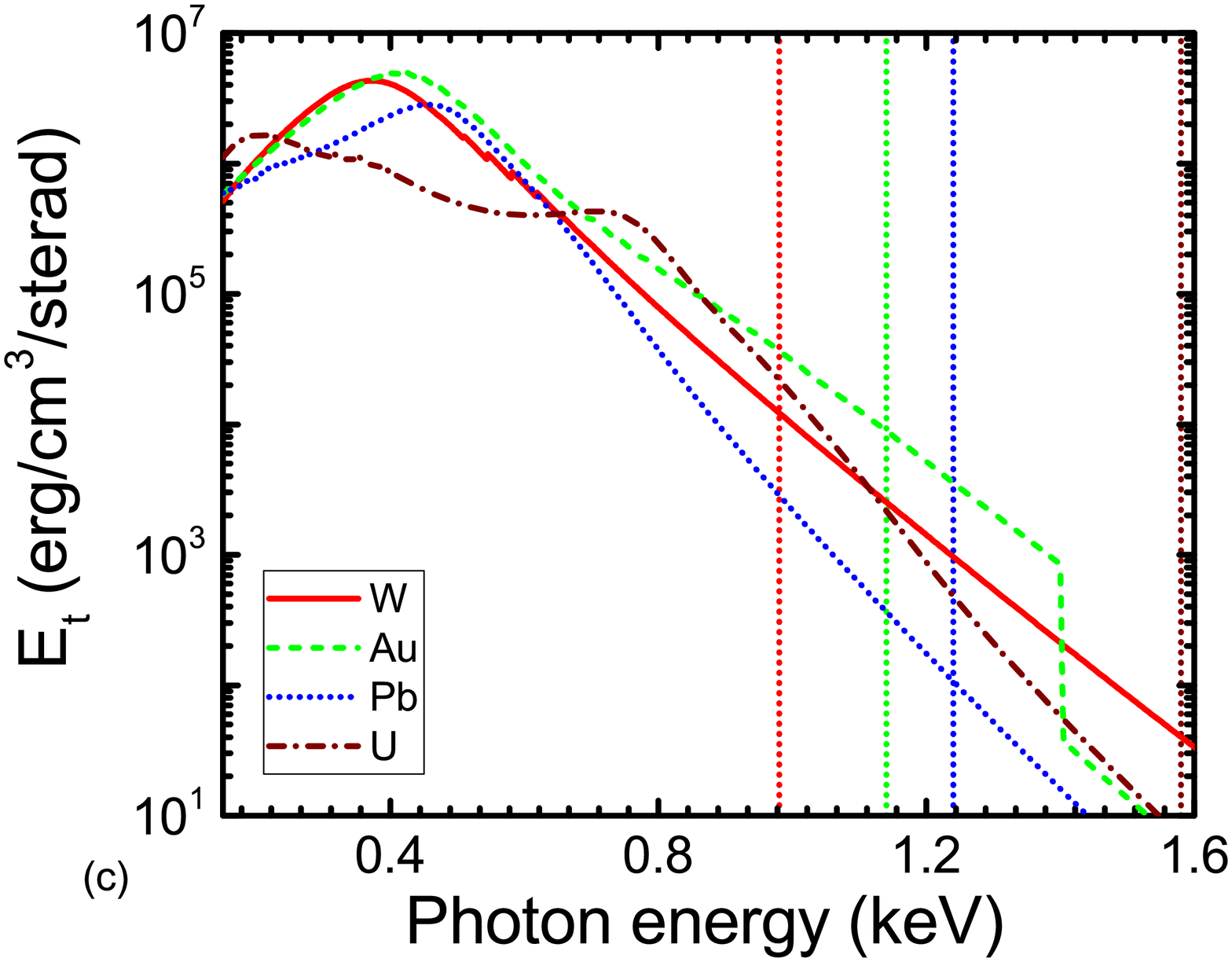}
   \caption{(Colour online only) (a) Radiation spectrum and emission coefficient evaluated at ($\rho$,$T_{e}$) of (b) conversion layer and (c) reemission zone against photon energy range 0.15-1.6 keV for different high-Z materials irradiated by laser intensity of $10^{14}$ W/cm\tsu{2}. The different vertical lines represent photon cut-off energies.\label{FS1e14-1}}
\end{figure}

\begin{figure}
        \includegraphics[width=0.49\textwidth]{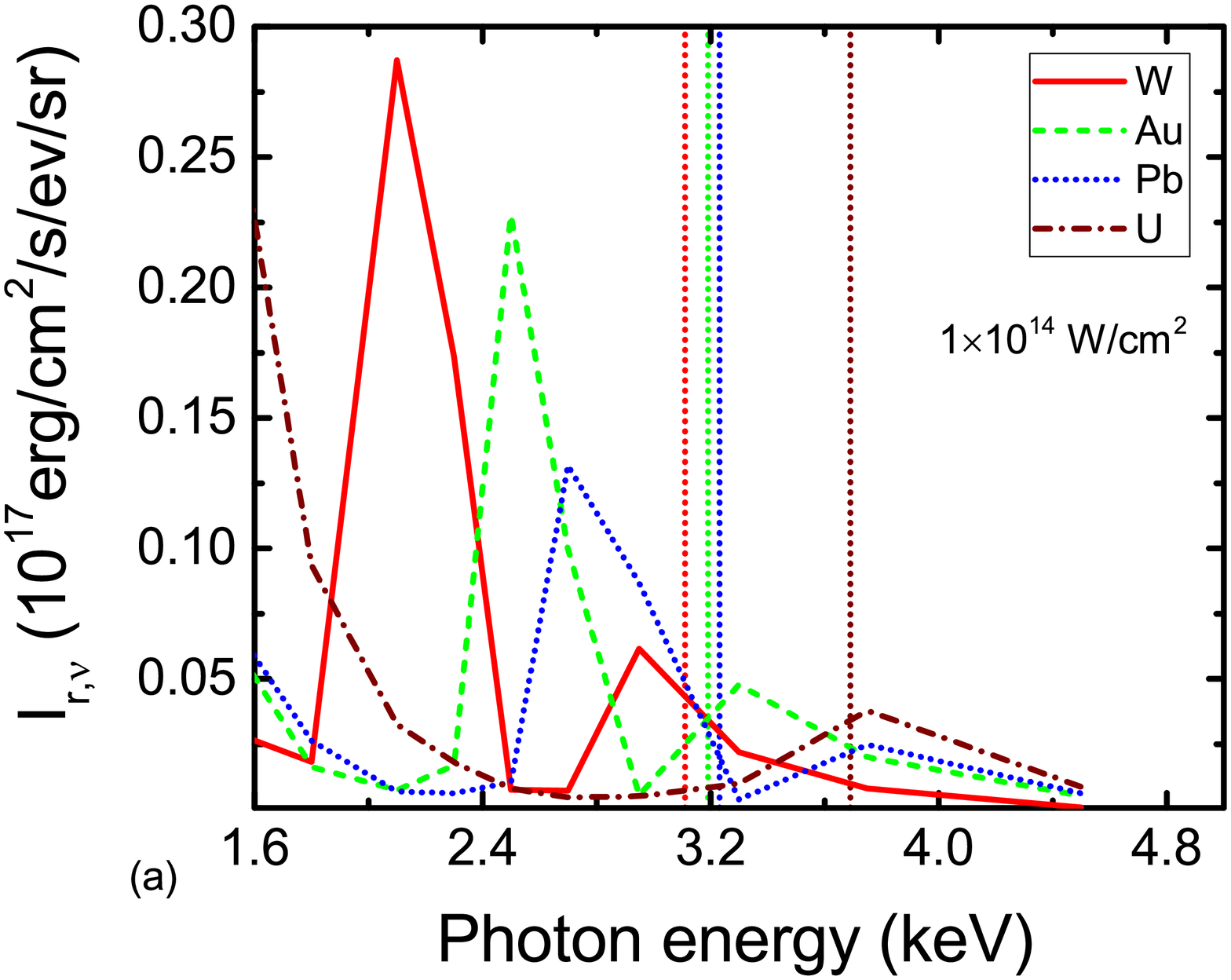}
        \includegraphics[width=0.49\textwidth]{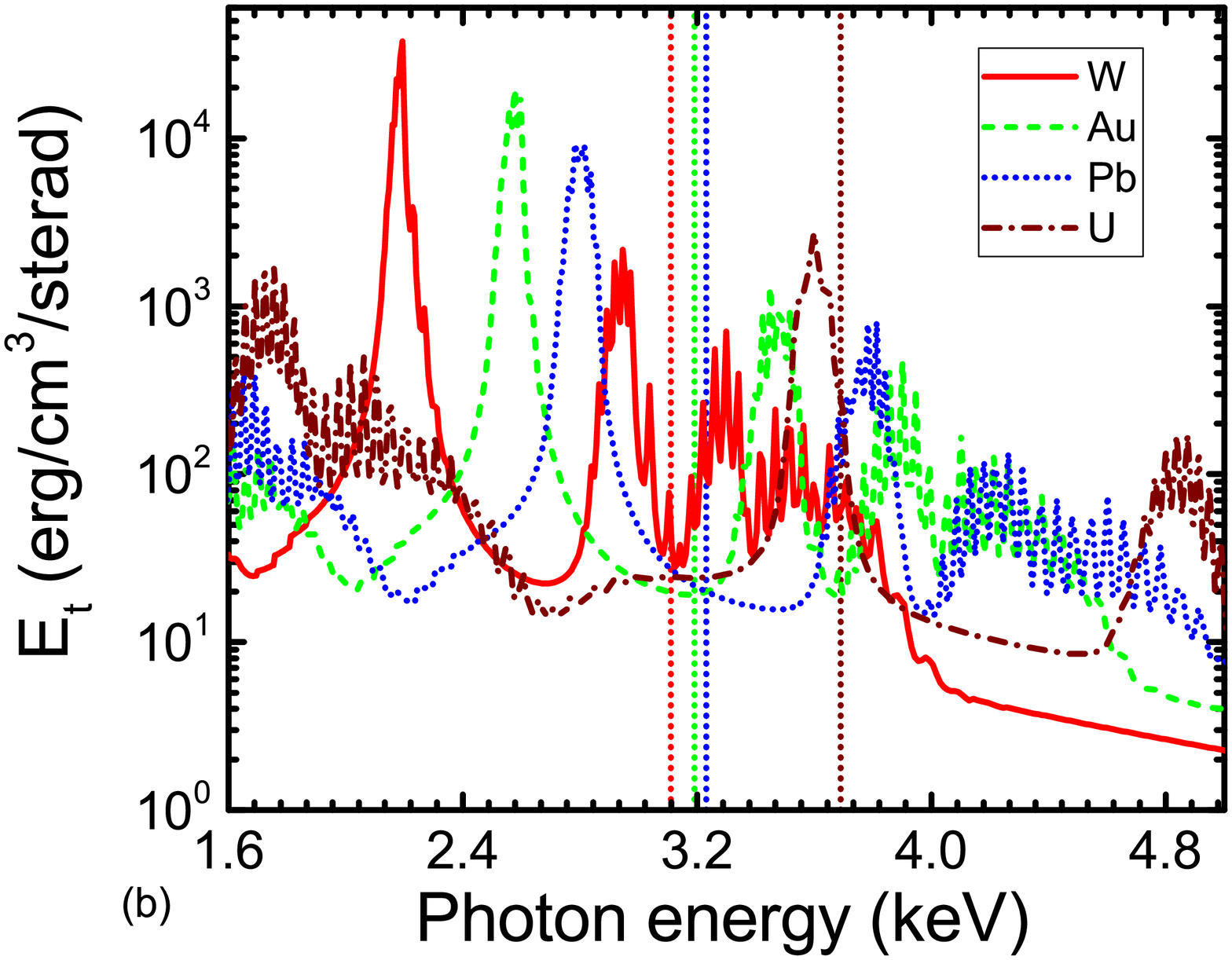}
\caption{(Colour online only) (a) Radiation spectrum and (b) emission coefficients evaluated at ($\rho$,$T_{e}$) of conversion layer against photon energy range 1.6-5.0 keV for different high-Z materials irradiated by laser intensity of $10^{14}$ W/cm\tsu{2}. The different vertical lines represent photon cut-off energies. \label{FS1e14-2}}
\end{figure}

Till now, we have discussed in detail the role of CL and RZ on conversion efficiency of Au foil and proposed a smooth broken power law for its variation with laser intensity. In the same way, x-ray conversion efficiencies are obtained in the other high-Z foils and a detailed comparative study is performed. The results of zoning analysis for the rest of the elements (W, Pb and U) are presented in Figs. \ref{AllEta}(a), \ref{AllEta}(c) and \ref{AllEta}(e).
The value of zone parameter for which $\eta$ is minimum, is found to be 0.96 for W and U whereas it is obtained as 0.97 for Pb. Here, we note that the value of optimum zone parameter remains nearly nondependent on the choice of target material under investigation. As explained earlier, usage of zone parameter is motivated to finely resolve the temperature and density gradients occurring close to conversion layer towards the laser irradiating side. In Figs. \ref{TempDensAll-0.6ns} (a) and (b), we have shown temperature and density variation against mass coordinate for all materials at their corresponding optimum zone parameters in region close to the conversion layer. All results correspond to time instant of 0.6 ns for a representative intensity of $4\times 10^{14}$ W/cm\tsu{2}. In each plot, left-bottom layer contains variations in W, Au and U whereas right-top layer consists of variation in Pb due to the differences in their mass densities. We observe similar gradients in temperature and density profiles for all materials, leading to nearly independent behaviour of zone parameter on the choice of different target materials. The simulated data for conversion efficiency and fitted curves (using Eq. \ref{EqEtaFit}) are shown for W, Pb and U in Figs. \ref{AllEta}(b), \ref{AllEta}(d) and \ref{AllEta}(f), respectively. The fitted results match well with simulated data for W and Pb but deviate a bit around break intensity for U. The disagreement observed around break intensity for U is due to poor selection of mesh width as observed from zoning analysis shown in Fig. \ref{AllEta} (e).
In the context of indirect drive ICF, soft x-rays are strongly absorbed by the low-Z ablator surrounding the fuel pellet thus causing uniform implosion of fuel. On the other hand, hard x-rays preheat the target thus decompressing the pellet that can compromise the final ignition. In our simulations, soft and hard (``M-band'') x-rays are defined in the photon energy ranges 0.15 keV to 1.6 keV (PR\tsb{1}) and  1.6 keV to 5 keV (PR\tsb{2}), respectively for all high-Z materials. For a better comparison, we have plotted the simulated data for soft ($\eta_S$) and M-band ($\eta_M$) x-ray conversion efficiencies  with laser intensity in Figs. \ref{EtaSMall} (a) and \ref{EtaSMall} (b), respectively.
We observe that $\eta_S$ attains maximum and minimum values for Au and U, respectively up to intensity of $\sim 3\times10^{13}$ W/cm\tsu{2} whereas that lies in between for Pb and W. On further increase in intensity, $\eta_S$ approaches nearly same values for Au and U whereas the value is minimum for W. On the other hand, different elements start showing significant contribution in $\eta_M$ for intensities higher than  $\sim 2\times 10^{13}$ W/cm\tsu{2}. We also note that W and U attain maximum and minimum $\eta_M$ with intermediate values realised by Au and Pb. To explain these results, we have considered two intensities  : one lower ($10^{13}$ W/cm\tsu{2}) and the other higher ($10^{14}$ W/cm\tsu{2}) than the intensity at which the conversion efficiency peaks. Fig.\ref{TRwithMassI} shows the plotted results of density ($\rho$), electron temperature ($T_e$) and radiation temperature ($T_{r}$) against mass coordinate ($m$) for a representative material Au at 0.6 ns. 
The plots depicted by red and blue colour correspond to lower and higher laser intensities, respectively. At the laser intensity of $10^{14}$ W/cm\tsu{2}, we observe that CL is fully developed with larger electron temperature and smaller density values compared to those at $10^{13}$ W/cm\tsu{2}. Further, CL shows a huge difference between $T_{e}$ and $T_{r}$, indicative of presence of strong NLTE conditions. On the other hand, similar values of $T_{e}$ and $T_{r}$ in RZ confirm LTE plasma conditions. The representative locations of RZ and CL are shown in Fig.\ref{TRwithMassI} by placing two vertical blue dotted lines in approximate midway of designated regions for the laser intensity of $10^{14}$ W/cm\tsu{2}. The corresponding values of $m$, $\rho$ and $T_e$ are given in Table \ref{table:rtHighI}. 
\begin{table*}
\caption{Mass coordinate, density and electron temperature  of representative locations describing reemission zone and conversion layer for different high-Z materials at laser intensity of $10^{14}$ W/cm\tsu{2}.\label{table:rtHighI}}
\begin{ruledtabular}
\begin{tabular}{ccccccccc}
                      &\multicolumn{2}{c}{W}& \multicolumn{2}{c}{Au}& \multicolumn{2}{c}{Pb}& \multicolumn{2}{c}{U}\\
                      &RZ     &CL      &RZ    &CL     &RZ    &CL     &RZ    &CL  \\ \hline
$m$ (mg/cm\tsu{2})    &3.48   &3.80    &3.47  &3.81   &1.88  &2.22   &3.36  &3.78\\ 
$\rho$ (g/cm\tsu{3})  &0.49   &0.0039  &0.53  &0.004  &0.53  &0.004  &0.61  &0.004\\
$T_e$ (eV)            &107    &1399    &103   &1532   &100   &1400   &101   &1326
\end{tabular}
\end{ruledtabular}
\end{table*}    
At laser intensity of $10^{13}$ W/cm\tsu{2}, CL is not fully developed with a relatively smaller difference between $T_{e}$ and $T_{r}$ compared to those at $10^{14}$ W/cm\tsu{2}, indicative of weak NLTE conditions (refer to Fig.\ref{TRwithMassI}). It may be noted that RZ still maintains LTE plasma conditions as shown by similar values of $T_e$ and $T_{r}$. As mentioned earlier, RZ is illustrated by placing red dotted vertical line in Fig.\ref{TRwithMassI}. We have not displayed the location of CL at this intensity as it overlaps with that attained at higher intensity. The representative locations of RZ and CL with corresponding density and temperature are given in Table \ref{table:rtLowI} at laser intensity of $10^{13}$ W/cm\tsu{2}.
\begin{table*}
\caption{Mass coordinate, density and electron temperature  of representative locations describing reemission zone and conversion layer for different high-Z materials at laser intensity of $10^{13}$ W/cm\tsu{2}.\label{table:rtLowI}}
\begin{ruledtabular}
\begin{tabular}{ccccccccc}
                      &\multicolumn{2}{c}{W}& \multicolumn{2}{c}{Au}& \multicolumn{2}{c}{Pb}& \multicolumn{2}{c}{U}\\
                      &RZ     &CL      &RZ    &CL     &RZ    &CL     &RZ    &CL  \\ \hline
$m$ (mg/cm\tsu{2})    &3.62   &3.80    &3.57  &3.81   &2.03  &2.22   &3.58  &3.78\\ 
$\rho$ (g/cm\tsu{3})  &0.38   &0.02    &0.42  &0.03   &0.40  &0.02   &0.38  &0.01\\
$T_e$ (eV)            &42     &133     &53    &125    &50    &172    &53    &244
\end{tabular}
\end{ruledtabular}
\end{table*}    
Rest of the high-Z elements also depict analogous plasma characteristics in RZ and CL at higher and lower laser intensities. With the help of Tables \ref{table:rtHighI} and \ref{table:rtLowI}, we are able to ascertain the hydrodynamic variables ($\rho$,$T_{e}$) realized in RZ and CL at two laser intensities for all high-Z elements. The set of ($\rho$,$T_{e}$) can be further used to determine the emission coefficient ($E_t$) variation with photon energy ($PE$) from SNOP opacity code for all high-Z elements in corresponding RZ and CL. In addition to this, we have to also define the cut-off photon energy in $E_{t}$ \emph{vs.} $PE$ plot for all high-Z elements beyond which the contribution of $E_{t}$ is not strong enough. This is accomplished by plotting the front spectrum ($I_{r,\nu}$) of x-rays for all elements at lower intensity in Fig. \ref{FS1e13-1} (a). 
As observed earlier, the hard x-ray contribution is negligible below 2\texttimes10\tsu{13} W/cm\tsu{2}, we have only plotted the front spectrum in range of $0.15-1.6$ keV. The cutoff photon energy ($PE_c$) is decided by including all important peaks  and ignoring contribution below 15\% of peak of front spectrum ($I{^{p}_{r,\nu}}$). For example, values of $I{^{p}_{r,\nu}}$, 0.15$I{^{p}_{r,\nu}}$ and $PE_c$ are obtained as $5.09 \times10^{16}$ erg/cm\tsu{2}/s/eV/sr, $0.764\times10^{16}$ erg/cm\tsu{2}/s/eV/sr and $1.04$ keV for Au at $10^{13}$ W/cm\tsu{2}. The locations of $PE_c$ thus obtained are depicted in Fig. \ref{FS1e13-1} (a) by placing dotted vertical lines for different elements. For clarity, we have also mentioned the values of $PE_c$ in Table \ref{table:PE_C} for all elements. 
\begin{table*}
\caption{Cut-off photon energies for various materials in different energy ranges ($PR_1=0.15-1.6$ keV and $PR_2=1.6-5$ keV)  at laser intensities of $10^{13}$ W/cm\tsu{2} and $10^{14}$ W/cm\tsu{2}.\label{table:PE_C}}
\begin{ruledtabular}
\begin{tabular}{ccccc}
                      &W &Au &Pb &U\\ \hline
$PE_c$ (keV)\footnotemark[1]    &0.6   &1.04    &0.64  &0.48\\ 
$PE_c$ (keV)\footnotemark[2]    &0.98  &1.14    &1.24  &1.58\\
$PE_c$ (keV)\footnotemark[3]    &3.11  &3.19    &3.23  &3.69
\end{tabular}
\end{ruledtabular}
\footnotetext[1]{$I_{L}$ = 10\tsu{13} W/cm\tsu{2}, $PR = PR_1$}
\footnotetext[2]{$I_{L}$ = 10\tsu{14} W/cm\tsu{2}, $PR = PR_1$}
\footnotetext[3]{$I_{L}$ = 10\tsu{14} W/cm\tsu{2}, $PR = PR_2$}
\end{table*}    
With the help of Tables \ref{table:rtLowI} and \ref{table:PE_C}, the plots of $E_{t}$ \emph{vs.} $PE$ with appropriate values of $PE_c$ are shown in Figs. \ref{FS1e13-1} (b) and (c) at corresponding $\rho$,$T_{e}$ values encountered in CL and RZ, respectively, for all elements. As observed earlier in Fig. \ref{EtaSMall}(a), $\eta_S$ reduces from Au to U while intermediate values are attained by W and Pb at laser intensity of $10^{13}$ W/cm\tsu{2}. Same trend is observed when contributions of $E_{t}$ along with corresponding $PE_c$ are taken into account for both CL and RZ regions.  For example, if we consider Au and U for comparison, we note that $E_{t}$ values for Au are always higher than that of U upto $PE_c$ of Au. This trend is only violated at a photon energy of 0.8 keV where U shows a peak. But this peak does not contribute significantly in x-ray emission as $E_t$ contributions for U are only effective upto the corresponding $PE_c$. On the other hand, contributions in $E_{t}$ are higher for U compared to that of Au for photon energies smaller than 0.279 keV in RZ. But the trend reverses after this $PE$ and Au shows dominant contributions compared to U up to $PE_c$ of U. Thus, we can explain the higher soft x-ray emission of Au compared to that of U by combining the results of both CL and RZ.  Due to under developed CL at lower laser intensities, we observe negligible M-band emission observed for all high Z elements below $2\times 10^{13}$ W/cm\tsu{2}. To explain the behaviour of soft x-ray emission at higher laser intensity, we have plotted the  front spectrum in Fig.\ref{FS1e14-1} (a) for photon energy range of 0.15 keV to 1.6 keV.
As mentioned earlier, the $PE_c$'s are determined by ignoring the contributions of $I_{r,\nu}$ beyond 0.15$I{^{p}_{r,\nu}}$ and the values are shown in Table \ref{table:PE_C}. With the help of Table \ref{table:rtHighI}, the emission coefficients are plotted at the corresponding densities and temperatures realized in CL and RZ of different materials in Figs. \ref{FS1e14-1} (b) and (c), respectively. The photon cut-off energies (refer to Table \ref{table:PE_C}) are also shown by dotted vertical lines. As observed earlier, Au and U show comparable soft x-ray emission with W showing the minimum at $10^{14}$ W/cm\tsu{2}. This trend is also noticed while considering the contribution of emission coefficients with appropriate cut-off photon energies from both CL and RZ. To understand the behaviour of different materials in terms of hard x-ray emission, we have plotted the front spectrum in Fig.\ref{FS1e14-2} (a) for $PE$ range of 1.6-5.0 keV at laser intensity of $10^{14}$ W/cm\tsu{2}.
The locations of $PE_c$ are also shown by vertical dotted lines and values are given in Table \ref{table:PE_C}. We have plotted the emission coefficients in Fig. \ref{FS1e14-2} (b) with corresponding $PE_c$ values at ($\rho$,$T_{e}$) (refer to Table \ref{table:rtHighI}) attained in CL. As observed in Fig.\ref{EtaSMall} (b), hard x-ray emission reduces from W to U with intermediate values realized by Au and Pb at higher intensities. This can be explained by considering the contributions of emission coefficients of different elements with appropriate $PE_{c}$ values as shown in Fig.\ref{FS1e14-2} (b). 
\section{Conclusion}\label{conclu}
In this paper, we have studied the phenomenon of x-ray emission in detail for tungsten, gold, lead and uranium planar foils driven by 1 ns flat top laser pulses by performing extensive RHD simulations. The  advantage of NLTE over LTE atomic physics is examined for the representative material gold. For NLTE opacities, a suitable  value of dielectronic recombination parameter is obtained by comparing the mean ionization state with other refined opacity models and experimental results. The choice of photon energy range, dielectronic recombination and zone parameter is found to be extremely important in obtaining accurate x-ray conversion efficiencies of laser driven gold foils. These appropriately chosen parameters are further used to evaluate the conversion efficiency of the four high-Z materials irradiated by 1ns flat top laser pulses with intensities varying in a wide range of $10^{12}-10^{15}$ W/cm\tsu{2}. A thorough zoning analysis for each laser intensity used for all materials illustrated the importance of non-uniform meshing instead of uniform meshing for laser generated plasma region. The variation of $\eta$ with laser intensity showed an optimum for all materials. Usage of LTE opacity in simulation significantly increased the conversion efficiencies along with the shift of optimum value towards higher intensity. The presence of optimum intensity is explained by numerically separating the laser plasma region in conversion layer and reemission zone. At any instant of time, total conversion efficiency is found to be derivable from characteristics emission contributions ($\eta{^{CL}_t}$ and $\alpha{^{RZ}_t}$) from CL and RZ. We observed RZ to follow LTE conditions but at higher intensity and later times, RZ is not able to contain M-band radiation leading to reduction in $\alpha^{RZ}_t$. In CL, $\eta^{CL}_t$ followed the same trend as $\eta_t$ with an optimum intensity that is an indicative of change in plasma condition from LTE to NLTE. Instead of separate scaling relations for different regions, we have proposed a generalized single scaling relation based upon smooth broken power law for conversion efficiency variation of different elements with laser intensity. The relationship is quite useful in strength characterization of various LPP sources as it directly connects the conversion efficiency of different elements with experimentally controllable and direct parameter - laser intensity. All of the materials are also explored for complete laser intensity range in terms of soft and M-band  x-ray conversion efficiencies. Up to a laser intensity of $10^{13}$ W/cm\tsu{2}, negligible M-band emission occurs and Au is the most preferable material in terms of $\eta_S$. However, at higher laser intensities (beyond $3 \times 10^{13}$ W/cm\tsu{2} ), performance of U is comparable or even better than Au, Pb and W due to maximum $\eta_S$ and minimum $\eta_M$ values. We have explained the results for different materials on the basis of emission coefficient contributions in both CL and RZ up to corresponding photon cut-off energies at different laser intensities. This analysis of x-ray emission in terms of emission coefficient contribution from CL and RZ is quite unique. In future, we aim to extend this approach to different methods of x-ray emission enhancement and look out for the new techniques based upon this analysis to further increase the strength of different LPP sources. With the recent applicability of low density/foam targets in NIF experiments, efforts will also be carried out to explore for the possibility for a universal conversion efficiency scaling relation that also includes material density along with the existing parameters. 
\section*{AUTHOR DECLARATIONS}
\section*{Conflict of interest}
The authors have no conflicts to disclose.
\section*{Data Availability Statement}
The data that support the findings of this study are available from the corresponding author upon reasonable request.
%

\end{document}